\documentclass[twocolumn,preprintnumbers,aps,showpacs,superscriptaddress,prx]{revtex4-1}
\usepackage{bm}
\usepackage[pdftex]{graphicx}
\usepackage{latexsym}
\usepackage{epsfig}
\usepackage{amsmath,amssymb}
\usepackage{amsfonts}
\usepackage{palatino}
\usepackage[latin1]{inputenc}
\usepackage{epsf}
\usepackage{latexsym}
\usepackage{calc}
\usepackage{color}
\usepackage{shadow}
\usepackage{epsfig}
\usepackage{float}
\usepackage[all]{xy}
\usepackage[pdftex, pdfstartview = {FitH}]{hyperref}
\usepackage{csquotes}
\usepackage{afterpage}

\usepackage{braket}

\def\br{{\bf r}}
\def\k{{\bf k}}
\newcommand{\sign}{\text{sign}}

\begin{document}

\title{Pseudospin anisotropy of trilayer semiconductor quantum Hall ferromagnets}

\date{\today}

\author{D. Miravet}
\email{dmiravet@gmail.com}
\author{C. R. Proetto}
\email{proetto@cab.cnea.gov.ar}
\affiliation{Centro At\'omico Bariloche and Instituto Balseiro, 8400 S. C. de Bariloche, R\'io Negro, Argentina}

\begin{abstract}
 When two Landau levels are brought to a close coincidence between them and with the chemical potential in the 
 Integer Quantum Hall regime, the two Landau levels can just cross or collapse while the external or
 pseudospin field that induces the alignment changes. In this work, all possible crossings are analyzed 
 theoretically for the particular case of semiconductor trilayer systems, using a variational Hartree-Fock approximation.
 The model includes tunneling between neighboring layers, bias, intra-layer and inter-layer
 Coulomb interaction among the electrons.
 We have found that the general pseudospin anisotropy classification scheme used in bilayers applies also to the trilayer situation,
 with the simple crossing corresponding to an easy-axis ferromagnetic anisotropy analogy, and the collapse case
 corresponding to an easy-plane ferromagnetic analogy. An isotropic case is also possible, with the levels just crossing or collapsing
 depending on the filling factor and the quantum numbers of the two nearby levels. While our results are valid for any integer filling factor $\nu$ (=1,2,3,...), we have analyzed in detail
 the crossings at $\nu=3$ and $4$, and we have given clear predictions that will help in their experimental search.
 In particular, the present calculations suggest that by increasing the bias, the trilayer system at these two filling factors can be driven from
 an easy-plane anisotropy regime to an easy-axis regime, and then can be driven back to the easy-plane regime.
 This kind of reentrant behavior is an unique feature of the trilayers, compared with the bilayers.
\end{abstract}

\maketitle

\maketitle

\section{Introduction}
The Integer Quantum Hall (IQH) effect in a semiconductor quasi two-dimensional electron gas (2DEG) is essentially a single-particle
phenomena~\cite{K05}. The magnetic field applied in the direction perpendicular to the layer quantizes the in-plane kinetic energy, forming
the celebrated Landau levels. The degeneracy of each of these Landau levels (LLs) is given by $N_{\phi}=AB/\Phi_0$, with
$A$ being the area of the 2DEG in the $x-y$ plane, $B$ is the magnetic field strength along the $z$ direction, and
$\Phi_0=ch/e$ is the magnetic flux number. The adimensional filling factor $\nu$ is defined as $\nu=N/N_{\phi}$, with $N$ being the 
total number of electrons; for 2DEG's, it is usually expressed as $\nu=(N/A)/(B/\Phi_0)$, in terms of 
the two-dimensional density $N/A$ of the 2DEG. For typical densities $N/A \sim 10^{11}/cm^2$ and $B ~ \sim$ some Teslas,
filling factors $\nu=1,2,3,...$ are easily achieved. At each and around of these integer filling factors, the 
chemical potential lies in the gap between two Landau levels, and the 2DEG behaves non-trivially, with zero
longitudinal and  Hall quantized resistances. These are the hallmarks of the IQH effect, first observed
by von Klitzing, Dorda and Pepper in 1980~\cite{vKDP80}. 

At even stronger magnetic fields, the increasing degeneracy of the Landau levels leads to filling factors smaller
than one, and the 2DEG enters in the Fractional Quantum Hall (FQH) regime, with a somehow similar experimental
phenomenology as in the integer case, but at particular fractional filling 
factors~\cite{TSG82}.
The stability of the 2DEG at these fractional $\nu$'s is understood as a many-body effect induced by the Coulomb
interaction among electrons in partially filled Landau levels~\cite{L83}.

However, many-body effects can also dominate the physics of the 2DEG even in the 
IQH regime, at the crossing of two Landau levels. The appearence of broken-symmetry states like easy-axis or 
easy-plane ferromagnets at these crossing situations has been termed under the name of Quantum Hall Ferromagnets 
(QHF)~\cite{Ezawa09}. Single-layer QHF have been extensively studied via magnetotransport 
measurements~\cite{Piazza99,Eom00,Smet01,Jaronszynski02}. Near LL crossings in tilted magnetic fields, the longitudinal
resistivity as a function of the in-plane component of the magnetic field exhibits hysteretic spikes, which
signals towards quantum Hall ferromagnetism~\cite{PTPS00}. The resistance spikes and hysteretic transport properties
were discussed theoretically in Ref.~[\onlinecite{JMD01}], using a self-consistent RPA/Hartree-Fock theory. 
In bilayer systems or single layers with two-subbands, spin-split LLs from distintic subbands cross even without a
tilted magnetic field, as observed experimentally~\cite{MPSHRP96,MSHKSE99,MSH01} and discussed 
theoretically~\cite{JSSSmD98,mDRJ99,BF00,JMD00,SmD00} in many previous works.
In particular, and based in a variational Hartree-Foch theory, Ref.~[\onlinecite{JMD00}] provides an exhaustive classification of the 
possible ferromagnetic anisotropies: depending on the quantum numbers of the two crossing levels, it was found that 
bilayers may exhibit isotropic, easy-plane, or easy-axis ferromagnetic states.
More recently, it was shown that near opposite spin LL crossings, magnetotransport and NMR measurements suggest a high
degree of spin polarization, which in turn points to a ferromagnetic instability of the 2DEG~\cite{ZMJ06}. Later
experiments using tilted magnetic fields lends further support to this suggestion~\cite{Guo08}. On the theoretical
side, in Ref.~[\onlinecite{Hao09}] the work of Jungwirth and MacDonald~\cite{JMD00} was generalized by allowing a
finite-width to the subband wavefunctions along the growth direction, the subband wavefunctions itself being
obtained through a self-consistent Local Density Approximation (LDA). The results were found in agreement with the 
experimental findings [\onlinecite{ZMJ06},\onlinecite{Guo08}], explaining the stability of the observed easy-plane (easy-axis) ground state at $\nu=3$ ($\nu=4$). 
It was also found a good agreement with the theoretical results of Ref.~(\cite{JMD00}), validating the strict-bidimensional approximation used in this work for 
the subband wavefunction in the growth direction.
Using spin density-functional theory,
plus a linear response approach for the magnetotransport calculations, Ref.~[\onlinecite{FFE10}] investigate the 
ring-like structures formed in the longitudinal resistivity $\rho_{xx}$ of a two-subband semiconductor quantum well, when plotted in a density-magnetic field
phase diagram~\cite{DGQLBP07}. Their theoretical findings are consistent with the experimental results~\cite{ZSJ07,Guo08} for the case of crossings
between opposite-spin LLs, but not for the case of crossings between same-spin LLs. For this later case, the authors suggested that a better
treatment than the local spin density approximation (LSDA) for the exchange functional may be needed~\cite{RP07}.

It is the aim of this work to discuss how many-body effects also manifest  in trilayer systems in the IQH regime
at the crossing between two Landau levels of the trilayer. As we will see, the three possible types of
magnetic anisotropies that are present in bilayers (isotropic, easy-axis, easy-plane) are also present  in trilayers,
but with some unique features that make these systems particularly attractive for experimental search.
For instance, we have found that by increasing the bias applied to the trilayers, the system can display easy-plane
anisotropy, then easy-axis anisotropy, and then again easy-plane anisotropy, both for the $\nu$ = 3 and 4 cases, with
the effect being however much more pronounced in the case $\nu=4$.
A similar reentrant behavior has been also found  at zero bias at $\nu=3$, with the energy of the central layer playing
a  similar role to the bias.

Trilayer systems have been already studied  theoretically at zero magnetic field~\cite{HHDV00,HP01,MPB16}, 
and in the IQH effect regime~\cite{HMD96}. This last work uses a theoretical approach similar to the one used here,
but the analysis is restricted to the case of one ($\nu=1$) or two ($\nu=2$) fully occupied Landau levels, and focused in 
the possibility that the trilayer system develops spontaneous interlayer coherence, even in the absence of
tunneling between layers, at these two filling factors. Here, we concentrate instead in the trilayer physics at the Landau
level crossings, and made clear predictions possible of be tested experimentally for the particular filling
factors $\nu=3$ and 4. On the experimental side, earlier publications report  the finding of IQH and FQH signals
in trilayers~\cite{JSESS92}, and  the evidence of a trilayer $\rightarrow$ bilayer transition induced by increasing
the perpendicular component of the magnetic field~\cite{SSS98,Comment,Reply}. More recent experimental work in trilayers 
concentrates on the multisubband fingerprints in the magnetoresistance oscillations as measured in Shubnikov-de Haas 
experiments at $\nu=2$ and 4~\cite{WMGRBP09}, 
on the effect of a tilted magnetic field on the IQH plateaus~\cite{GWRBP09}, and report spectroscopic 
evidence on the collapse of the interlayer
tunneling gap for particular values of the tilted component of the magnetic field~\cite{dS14}.
The unique features of the IQH effect in trilayer graphene has been also studied~\cite{TWTJ11}.
The authors observed at high magnetic fields, that the degenerate crossing points splits into manifolds,
and suggested from this the existence of broken-symmetry quantum Hall states. 

The rest of the paper is organized as follows: In Sec.~II we explain the model, introduce the corresponding
single-particle states and explain the pseudospin analogy for its representation. Sec.~III is devoted to the 
building of the many-body Hamiltonian in the restricted pseudospin subspace of the two crossing Landau levels,
and how its solution is obtained within a variational Hartree-Fock method. In Sec.~IV we analyze in detail the magnetic anisotropy
terms which give rise to the easy-axis, the easy-plane and the isotropic classification scheme for the crossings, while
in Sec.~V the effect of the finite pseudospin field is determined. Finally, the Conclusions are given in Sec.~VI. 

\section{Model and single-particle states}
Let us start from the simplest possible case: the trilayer at zero magnetic field~\cite{HHDV00,HP01,MPB16}.
In the absence of a magnetic field perpendicular to the layers, and assuming translational invariance along the 
layers, the electronic single-particle states may be written in the factorized form
\begin{equation}
 \psi_{\xi \, \k \, \sigma}(\br) = \frac{e^{i \k \cdot \boldsymbol\rho}}{A} \lambda_{\xi}(z) \, \eta_{\sigma} \; , 
 \label{eq:zero-field}
\end{equation}
where $\br=(\boldsymbol\rho,z)$ with $\boldsymbol\rho$ the in-plane coordinate, $\xi$ is the subband index, 
and $\k = (k_x,k_y)$ is the in-plane wave vector. 
$\sigma$ is the spin index, which can take the values $\pm$ $\frac{1}{2}$.
and $\eta_{\sigma}$ is the spin-1/2 spinor, such that

\begin{equation}
\eta_+ = \begin{bmatrix}
1 \\
0
\end{bmatrix}
\qquad , \,
\eta_- = \begin{bmatrix}
0 \\
1
\end{bmatrix}
\; .
\end{equation}

Within our simple model for the trilayer, as schematized in Fig.~1, 
the normalized subband wave functions $\lambda_{\xi}(z)$
are written as
\begin{equation}
 \lambda_{\xi}(z) = a_{\xi} \sqrt{\delta\left(z+d\right)}
                            + b_{\xi} \sqrt{\delta\left(z\right)}
                            + c_{\xi} \sqrt{\delta\left(z-d\right)} \; ,
 \label{eq:subband WF}
\end{equation}
with $\xi=-1,0,1$, and the coefficients $a_{\xi},b_{\xi},c_{\xi}$ 
being the eigenstates of the 3 $\times$ 3 tight-binding matrix
\begin{equation}
\left(\begin{array}{ccc}
\varepsilon_{1} & -t & 0\\
-t & \varepsilon_{2} & -t\\
0 & -t & \varepsilon_{3}
\end{array}\right)
\label{eq:tigth_binding_matrix}
\end{equation}
$\varepsilon_1$, $\varepsilon_2$, and $\varepsilon_3$ are the diagonal energies for electrons in layers 1, 2, and 3, respectively.
In our model, each of the three 2DEG's is represented by an strictly bidimensional metallic layer; this is the 
approximation behind Eq.~(\ref{eq:subband WF}). However, this approximation is not essential and can be relaxed as it has been already done in the case of bilayers at zero~\cite{RP97} and finite~\cite{Hao09} magnetic fields. These more elaborated calculations,
where the wavefunctions $\lambda_{\xi}(z)$ have a finite width at the layers, shows only quantitative
differences in comparison with the strict 2D approximation. We expect that the same type of considerations 
regarding this issue will apply to our trilayer model.

Calling $\gamma_{\xi}(\varepsilon_1,\varepsilon_2,\varepsilon_3,t)$ to the corresponding eigenvalues of Eq.~(\ref{eq:tigth_binding_matrix})
($\gamma_{-1} \le \gamma_{0} \le \gamma_{1}$), the solutions associated with the wave functions
in Eq.~(\ref{eq:zero-field}) have the zero-field energies $E_{\xi}(\k)= \gamma_{\xi} + \hbar^2 k^2 /(2 m^*)$.
$m^*$ is the effective mass for electrons in the well-acting semiconductor, typically GaAs. 
Due to the in-plane kinetic energy, the zero-field energy spectrum is continuous. 

\begin{figure}[h]
\begin{center}
\includegraphics[width=8.6cm]{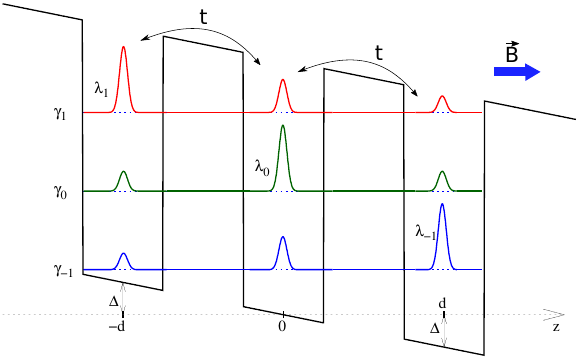}
\caption{Schematic view of the trilayer system. $\gamma_{-1} \; (\lambda_{-1}(z))$, $\gamma_{0} \; (\lambda_{0}(z))$ 
and $\gamma_{1} \; (\lambda_{1}(z))$ are the eigenvalues
(eigenvectors) of the $3 \times 3$ matrix in Eq.~(\ref{eq:tigth_binding_matrix}). $t$ and $\Delta$ represents the quantum
mechanical tunneling and bias between neighboring layers, respectively. $d$ is the distance between quantum-well centers, and
a magnetic field of amplitude $B$ is applied along the $z$ direction (thick arrow).}
\label{Fig1}
\end{center}
\end{figure}

The physical situation changes dramatically when a magnetic field is applied in the direction perpendicular
to the layers. In this case, the eigenstates still may be expressed in a factorized form, but the in-plane
factor must be replaced,
\begin{equation}
 \frac{e^{i \k \cdot \boldsymbol\rho}}{A} \rightarrow \phi_{n,k_y}(x) \frac{e^{iyk_y}}{\sqrt{L_y}} \;,
\end{equation}
with
\begin{equation}
 \phi_{n,k}(x) = \frac{\exp\left[-\frac{\left(x-l_B^{2}k\right)^{2}}{2l^{2}}\right]}
                 {\left[\sqrt{\pi} \, l_B \, 2^{n}\left(n!\right)\right]^{1/2}}
                 H_{n}\left(\frac{x-l_B^{2}k}{l_B}\right) \;.
\end{equation}
Here $H_n(x)$ are the $n$-th order Hermite polynomials, $n$ (= 0,1,2,...) is the Landau level orbital quantum number,
and $k$ $(= 1,2, ... , N_{\phi})$ is the one-dimensional wave vector label that distinguishes states within a given LL.
$l_B=\sqrt{c\hbar/eB}$ is the magnetic length. 
The single-particle energy spectrum consists now of discrete LLs,
\begin{equation}
 E_{\xi\,n\,\sigma}=\gamma_{\xi} + \left(n+1/2\right)\hbar\omega_{c}-\sigma \left|g\right|\mu_{B}B \;,
\end{equation}
where $\omega_c = eB/(m^*c)$ is the cyclotron frequency, and the last term is the real-spin Zeeman coupling.
Note that $E_{\xi\,n\,\sigma}$ does not depend on $k_y$, and this explains the macroscopic 
$N_{\phi}$ degeneracy of each LL, the same for all of them.

The Coulomb interaction among the electrons,
whose effects will be analyzed in detail in the following section, mixes (in principle) all the LLs, 
and then the single-particle labels $\xi,\,n,\,\sigma$ loose the nice property of being ``good quantum numbers''. 
In this work, we are interested in the possible many-body ground-states that may occur when two LLs are brought
close to alignment while remaining sufficiently separated from all other LLs. Following the strategy of
Ref.~(\onlinecite{JMD00}) for the bilayer case, we will simplify the full interacting problem described above by
considering explicitly only the Coulomb-induced mixing between the two LLs close to degeneracy at chemical potential, while including
the effect of the totally filled lower energy LLs in the form of single-particle effective fields. We will denote with the 
symbol $p=\uparrow,\downarrow$ to each of the two approaching LLs, and we will refer to it as a {\it pseudospin} 
index or label. The two close to alignment LLs will have single-particle quantum numbers 
$\xi(\uparrow),n(\uparrow),\sigma(\uparrow)$ and $\xi(\downarrow),n(\downarrow),\sigma(\downarrow)$ respectively,
and at least one of them should be different.

Within this truncated model, the full set of single-particle states reduces then to the following wave functions,
\begin{equation}
 \psi_{p,k_y}(\br)=
 \lambda_{\xi(p)}(z) \phi_{n(p),\sigma(p),k_y}(x)\frac{\exp(iyk_y)}{\sqrt{L_{y}}} \; ,
 \label{eq:single_pseudospin}
\end{equation}
where $\phi_{n(p),\sigma(p),k_y}(x) = \phi_{n(p),k_y}(x) \times \eta_{\sigma(p)}$.

In a second-quantization language, the operator that creates a particle in a state with a pseudospin
oriented in a arbitrary direction $\hat{m}=(\sin\theta\cos\varphi,\sin\theta\sin\varphi,\cos\theta)$
is described by 
\begin{equation}
 \hat{c}_{\hat{m},k_y}^{\dagger}=
                           \cos\left(\frac{\theta}{2}\right)\hat{c}_{\uparrow,k_y}^{\dagger}
                         + \sin\left(\frac{\theta}{2}\right)e^{i\varphi}\hat{c}_{\downarrow,k_y}^{\dagger} \;,
 \label{co}
\end{equation}
where $\hat{c}_{p,k}^{\dagger}$ creates a single-particle state whose wave function is given in 
Eq.~(\ref{eq:single_pseudospin}). As we will see, the value of $\hat{m}$ at crossing will allow us to provide a general classification scheme
for all possible coincidences in trilayers. Note that for $\theta=0 \, (m_z=+1)$ or $\theta=\pi \, (m_z=-1)$, the electron is in a pure pseudospin state $\uparrow$
or $\downarrow$,  while for any other value of $\theta$ the electron is in a mixed pseudospin state.

\section{Many-body Hamiltonian }

The many-body Hamiltonian that represents the interaction between two
crossing LLs, having into account the mean field contribution
of lower filled LLs, can be written in the truncated two-dimensional pseudospin Hilbert space in the compact form
\begin{eqnarray}
 \hat{H} &=& -\sum_{i=\mathbf{1},x,y,z}\sum_{k=1}^{N_{\phi}}\sum_{\alpha,\alpha'=1}^2 b_{i}\,\tau_{i}^{\alpha',\alpha}\,
 \hat{c}_{p(\alpha'),k}^{\dagger}\hat{c}_{p(\alpha),k} \nonumber \\
 &+& \frac{1}{2}\sum_{i,j=\mathbf{1},x,y,z}\sum_{\substack{k_{1},k_{1}' \\ k_{2},k_{2}'=1}}^{N_{\phi}}
 \sum_{\substack{\alpha_{1},\alpha_{1}' \\ \alpha_{2},\alpha_{2}'=1}}^{2}
 W_{i,j}^{k'_{1},k_{2}',k_{1},k_{2}} \\
 &\times& \tau_{i}^{\alpha_{1}',\alpha_{1}}\tau_{j}^{\alpha_{2}',\alpha{}_{2}} \,
 \hat{c}_{p(\alpha'_{1}),k'_{1}}^{\dagger}\hat{c}_{p(\alpha'_{2}),k_{2}'}^{\dagger}
 \hat{c}_{p(\alpha_{2}),k_{2}}\hat{c}_{p(\alpha_{1}),k_{1}} \nonumber \; ,
 \label{Hamiltonian}
\end{eqnarray}
where $p(1)=\, \uparrow$, $p(2)=\,\downarrow$, $\tau_{x}$, $\tau_{y}$, $\tau_{z}$ are the Pauli spin matrices,
and $\tau_{\mathbf{1}}$ is the $2\times2$ identity matrix. The potential
$W_{i,j}$ represents different combinations (see Eq.~(\ref{alfas}) below) of the Coulomb interaction
matrix elements $V_{p'_{1},p'_{2},p_{1},p_{2}}$

\begin{eqnarray}
 V_{p'_{1},p'_{2},p_{1},p_{2}}^{k'_{1},k_{2}',k_{1},k_{2}} &=& 
 \int d^{3}r_{1}\int d^{3}r_{2}~\psi^*_{p_{1}',k_{1}'}({\br}_{1})\psi^*_{p_{2}',k_{2}'}({\br}_{2}) \nonumber \\
 &\times& \frac{e^{2}}{\epsilon|{\br}_{1}-{\br}_{2}|}\psi_{p_{1},k_{1}}({\br}_{1})\psi_{p_{2},k_{2}}({\br}_{2}) \; .
 \label{V}
\end{eqnarray}
$\epsilon$ is the dielectric constant of the semiconductor well material ($\sim$ 12.5 for GaAs).
The single Slater approximation to the many-electron state with pseudospin orientation $\hat{m}$ is
expressed in the form
\begin{equation}
 \Ket{\psi\left[\hat{m}\right]}=\prod_{k=1}^{N_{\phi}}c_{\hat{m}k}^{\dagger}\Ket0 \; ,
 \label{HFstate}
\end{equation}
with the $\hat{c}_{\hat{m},k}^{\dagger}$ as defined in Eq.~(\ref{co}).
Then we find the Hartree-Fock  energy per particle as follows, 
\begin{eqnarray}
 e_{HF}(\hat{m}) &\equiv& \frac{\Bra{\psi\left[\hat{m}\right]}\hat{H}\Ket{\psi\left[\hat{m}\right]}}{N_{\phi}} \\
 &=& -\sum_{i=x,y,z}\left(b_{i}-\frac{1}{2}U_{\mathbf{1},i}-\frac{1}{2}U_{i,\mathbf{1}}\right)m_{i}\label{eq:HF-energy}
 \nonumber \\
 &+& \frac{1}{2}\sum_{i,j=x,y,z}U_{i,j}m_{i}m_{j} \; ,
 \label{HF_energy*}
\end{eqnarray}
 where 
\begin{equation}
 U_{i,j}=\frac{1}{N_{\phi}}\sum_{k_{1},k_{2}=1}^{N_{\Phi}}
         \left(W_{i,j}^{k_{1},k_{2},k_{1},k_{2}}-W_{i,j}^{k_{2},k_{1},k_{1}k_{2}}\right).
 \label{Uij*}        
\end{equation}

The quantities $W_{i,j}^{k_{1}',k_{2}',k_{1},k_{2}}$ in Eq.~(\ref{Uij*}) are related to the 
$V_{p'_{1},p'_{2},p_{1},p_{2}}^{k'_{1},k_{2}',k_{1},k_{2}}$ through the following definition,
\begin{equation}
 W_{i,j}^{k_{1}',k_{2}',k_{1},k_{2}} = \sum_{p_1',p_2',p_1,p_2} \alpha_{i,j}^{p_1',p_2',p_1,p_2} \;
 V_{p'_{1},p'_{2},p_{1},p_{2}}^{k'_{1},k_{2}',k_{1},k_{2}} \; ,
 \label{alfas}
\end{equation}
where the coefficients $\alpha_{i,j}^{p_1',p_2',p_1,p_2}$ are given in Table \ref{Table2}. 
Replacing this definition in Eq.~(\ref{Uij*}), yields
\begin{eqnarray}
 U_{i,j} &=& \frac{1}{N_{\phi}} \sum_{p_1',p_2',p_1,p_2} \alpha_{i,j}^{p_1',p_2',p_1,p_2}
         \nonumber \\
         &\times& \sum_{k_{1},k_{2}=1}^{N_{\Phi}} \left( V_{p'_{1},p'_{2},p_{1},p_{2}}^{k_{1},k_{2},k_{1},k_{2}} 
         -V_{p'_{1},p'_{2},p_{1},p_{2}}^{k_{2},k_{1},k_{1},k_{2}} \right) \nonumber \; , \\
         &=& \sum_{p_1',p_2',p_1,p_2} \alpha_{i,j}^{p_1',p_2',p_1,p_2} \nonumber \\
         &\times& \frac{1}{2\pi}\left(v_{p_{1}',p_{2}',p_{1},p_{2}}(0)-\int_{0}^{\infty} q e^{-q^{2}/2}v_{p_{1}',p_{2}',p_{1},p_{2}}({q})dq\right) . \nonumber \\
         \label{Uij}        
\end{eqnarray}
In passing from the first to the second line in above equation  we have used the quasi-2D Fourier representation of the Coulomb interaction,
for obtaining the relation~\cite{JMD00}
\begin{eqnarray}
 & \frac{1}{N_{\phi}}\sum\limits_{k_{1},k_{2}=1}^{N_{\phi}}
 \left(V_{p'_{1},p'_{2},p_{1},p_{2}}^{k_{1},k_{2},k_{1},k_{2}}
 -V_{p'_{1},p'_{2},p_{1},p_{2}}^{k_{2},k_{1},k_{1}k_{2}}\right)\nonumber \\
 = & \int\frac{d^{2}{q}}{(2\pi)^{2}}e^{-q^{2}/2}
 \left[v_{p_{1}',p_{2}',p_{1},p_{2}}(0)-v_{p_{1}',p_{2}',p_{1},p_{2}}(q)\right], \nonumber \\
 =& \frac{1}{2\pi}\left(v_{p_{1}',p_{2}',p_{1},p_{2}}(0)-\int_{0}^{\infty} q e^{-q^{2}/2}v_{p_{1}',p_{2}',p_{1},p_{2}}({q})dq\right) . \nonumber \\
 \label{eq:Hartre-Intra}\nonumber \\
\end{eqnarray}
$v_{p_{1}',p_{2}',p_{1},p_{2}}(q)$ is the product of two terms,
\begin{equation}
 v_{p_{1}',p_{2}',p_{1},p_{2}}({ q})=v_{p_{1}',p_{2}',p_{1},p_{2}}^{\parallel}({q})
 v_{\xi(p_{1}'),\xi(p_{2}'),\xi(p_{1}),\xi(p_{2})}^{\perp}({q}) \; .
 \label{v}
\end{equation}
$v_{p_{1}',p_{2}',p_{1},p_{2}}^{\perp}(q)$ is the
subband factor and $v_{p_{1}',p_{2}',p_{1},p_{2}}^{\parallel}(q)$
is the in-plane factor. This last factor depends only on the wave function
in 2D plane that is the same in bilayers and trilayers. For this reason
the only term that differs from the bilayer case is the subband factor.
In this last equation and  in the following of this work we have used $l_B$ (the magnetic length) as
unit of length and $e^{2}/\epsilon l_B$ as unit of energy. The first
factor in Eq.~(\ref{Uij}) at $q=0$ represents the Hartree contribution
and the second factor corresponds to the exchange contribution.

The subband factor is defined by
\begin{eqnarray}
 v_{\xi(p_{1}'),\xi(p_{2}'),\xi(p_{1}),\xi(p_{2})}^{\perp}({q}) = 
 \int_{-\infty}^{\infty}dz_{1}\int_{-\infty}^{\infty}dz_{2} & \; e^{-q|z_{1}-z_{2}|}  \nonumber \\
 \times \lambda_{\xi(p'_{1})}(z_{1})\lambda_{\xi(p_{2}')}(z_{2}) 
 \lambda_{\xi(p_{1})}(z_{1})\lambda_{\xi(p_{2})}(z_{2}) . & 
 \label{eq:Subband_factor_definition}
\end{eqnarray}

From Eqs. (\ref{eq:subband WF}) and (\ref{eq:Subband_factor_definition})
we have 

\begin{multline}
 v_{\xi(p_{1}'),\xi(p_{2}'),\xi(p_{1}),\xi(p_{2})}^{\perp}(q) = a_{\xi(p_{1}')}a_{\xi(p_{2}')}a_{\xi(p_{1})}a_{\xi(p_{2})}\\
 +b_{\xi(p_{1}')}b_{\xi(p_{2}')}b_{\xi(p_{1})}b_{\xi(p_{2})}+c_{\xi(p_{1}')}c_{\xi(p_{2}')}c_{\xi(p_{1})}c_{\xi(p_{2})}\\
 +\left[a_{\xi(p_{1}')}b_{\xi(p_{2}')}a_{\xi(p_{1})}b_{\xi(p_{2})}+b_{\xi(p_{1}')}a_{\xi(p_{2}')}b_{\xi(p_{1})}a_{\xi(p_{2})}\right]e^{-qd}\\
 +\left[b_{\xi(p_{1}')}c_{\xi(p_{2}')}b_{\xi(p_{1})}c_{\xi(p_{2})}+c_{\xi(p_{1}')}b_{\xi(p_{2}')}c_{\xi(p_{1})}b_{\xi(p_{2})}\right]e^{-qd}\\
 +\left[a_{\xi(p_{1}')}c_{\xi(p_{2}')}a_{\xi(p_{1})}c_{\xi(p_{2})}+c_{\xi(p_{1}')}a_{\xi(p_{2}')}c_{\xi(p_{1})}a_{\xi(p_{2})}\right]e^{-2qd}.
 \label{eq:subband factor_giant}
\end{multline}

Note that the expression (\ref{eq:subband factor_giant}) reproduces
all cases of a bilayer system if we take $b_{\xi}=0$ and $a_{\xi}$,
$c_{\xi}$ as given by the Eq.~(1) of Ref.~(\onlinecite{JMD00}). 

The in-plane factor is given by 
\begin{eqnarray}
 && \quad \qquad v_{p_{1}',p_{2}',p_{1},p_{2}}^{\parallel}(q) = 
 \frac{{e^{{q^2}/{2}}}}{q}\int d\Omega \nonumber \\
 &\times& \int\limits_{-\infty}^{\infty}dx_{1} \, e^{i q_x x_1} \phi_{n(p_{1}'),\sigma(p_1'),q_{y}/2}^{\dagger}(x_{1})
 \phi_{n(p_{1}),\sigma(p_1),-q_{y}/2}(x_{1}) \nonumber \\
 &\times& \int\limits_{-\infty}^{\infty}dx_{2} \, e^{-i q_x x_2}\phi_{n(p_{2}'),\sigma(p_2'),-q_{y}/2}^{\dagger}(x_{2})
 \phi_{n(p_{2}),\sigma(p_2),q_{y}/2}(x_{2}) \; , \nonumber \\
 &=& \delta_{\sigma(p_1'),\sigma(p_1)} \delta_{\sigma(p_2'),\sigma(p_2)}\delta_{n(p_1')-n(p_1),n(p_2)-n(p_2')}
 \left(\frac{n_1^<!\,n_2^<!}{n_1^>\,!n_2^>!}\right)^{1/2} \nonumber \\
 &\times& \frac{2\pi}{q} \left(\frac{q^2}{2}\right)^{n_1^>-n_1^<}
 L_{n_1^<}^{n_1^>-n_1^<}\left(\frac{q^2}{2}\right) L_{n_2^<}^{n_2^>-n_2^<}\left(\frac{q^2}{2}\right) \nonumber \; ,\\
 \label{master}
\end{eqnarray}
where $d\Omega$ is the 2D solid angle.
Here $L_{n}^m\left(x\right)$ are the generalized Laguerre polynomials, 
$n_i^< \equiv \min[n(p_i),n(p_{i}')]$ and $n_i^> \equiv \max[n(p_i),n(p_{i}')]$.
For later use, $L_n^{(0)}(x)=L_n(x)$ are the Laguerre polynomials.

Before concluding this section, it is important to emphasize the physical content of the variational Hartree-Fock state 
in Eq.~(\ref{HFstate}): it is a single Slater state for $N_{\phi}$ electrons, all having the same value of $\hat{m}$, 
and remembering that $N_{\phi}$ is the exact degeneracy of each LLs. As the angles $\theta$ and $\varphi$ are  not fixed,
they provide the minimizing energy parameters in all later calculations.
It is also worth at this point to remark that for all derivations in this section we have used the general expression for the wave
function $\psi_{p,k_y}(\br)$ in Eq.~(\ref{eq:single_pseudospin}); only in Eq.(\ref{eq:subband factor_giant}) we have used the particular expression of
Eq.~(\ref{eq:subband WF}) for the subband wavefunctions $\lambda_{\xi}(z)$. In particular, the crucial factorization in Eq.~(\ref{v}) is valid as long as the 
factorization in Eq.~(\ref{eq:single_pseudospin}) is fulfilled. The different approximations for $\lambda_{\xi}(z)$ will only impact on
$v_{\xi(p_{1}'),\xi(p_{2}'),\xi(p_{1}),\xi(p_{2})}^{\perp}(q)$, that in turn as we will see below may lend only to quantitative changes in the results.

\begin{table*}%%[h]
\begin{center}
\caption{Each entry in the table defines the coefficient $4\alpha_{ij}^{p_1',p_2',p_1,p_2}$. Only the four pseudospin
indices are shown as up and down arrows, and they should be multiplied by the factor $-i$ when indicated. As an example of the use of the table,
for $i=j=z$, the only non-zero coefficients are $\alpha_{zz}^{\uparrow \uparrow \uparrow \uparrow} = \alpha_{zz}^{\downarrow \downarrow \downarrow \downarrow} = 1/4$
and $\alpha_{zz}^{\uparrow \downarrow \uparrow \downarrow} = \alpha_{zz}^{\downarrow \uparrow \downarrow \uparrow} = -1/4$.}
\label{Table2}
\begin{tabular*}{0.993\textwidth}{|l|l|l|l|l|}

\hline
\hline

    &\multicolumn{1}{|c|}{$1$} & \multicolumn{1}{c|}{$x$} & \multicolumn{1}{c|}{$y$} & \multicolumn{1}{c|}{$z$}  \\
\hline
%%\hline
$1$ & $\uparrow\uparrow\uparrow\uparrow=\downarrow\downarrow\downarrow\downarrow=\uparrow\downarrow\uparrow\downarrow=\downarrow\uparrow\downarrow\uparrow$ & $\uparrow\uparrow\uparrow\downarrow=\uparrow\downarrow\uparrow\downarrow=\downarrow\uparrow\downarrow\uparrow=\downarrow\downarrow\downarrow\uparrow$ & $\uparrow\uparrow\uparrow\downarrow=-\uparrow\downarrow\uparrow\uparrow=\downarrow\uparrow\downarrow\uparrow=-\downarrow\downarrow\downarrow\uparrow$ & $\uparrow\uparrow\uparrow\uparrow=-\downarrow\downarrow\downarrow\downarrow=-\uparrow\downarrow\uparrow\downarrow=\downarrow\uparrow\downarrow\uparrow$ \\

   &   &   & \textcolor{red}{$(-i)$ } &   \\
\hline
$x$ & $\uparrow\uparrow\downarrow\uparrow=\downarrow\uparrow\uparrow\uparrow=\uparrow\downarrow\downarrow\downarrow=\downarrow\downarrow\uparrow\downarrow$ & $\uparrow\downarrow\downarrow\uparrow=\downarrow\uparrow\uparrow\downarrow=\uparrow\uparrow\downarrow\downarrow=\downarrow\downarrow\uparrow\uparrow$ & $\uparrow\downarrow\downarrow\uparrow=-\downarrow\uparrow\uparrow\downarrow=-\uparrow\uparrow\downarrow\downarrow=\downarrow\downarrow\uparrow\uparrow$ & $\uparrow\uparrow\downarrow\uparrow=\downarrow\uparrow\uparrow\uparrow=-\uparrow\downarrow\downarrow\downarrow=-\downarrow\downarrow\uparrow\downarrow$  \\

  &   &   &\textcolor{red}{$(-i)$ }   &   \\

\hline
$y$ & $\uparrow\uparrow\downarrow\uparrow=-\downarrow\uparrow\uparrow\downarrow=\uparrow\downarrow\downarrow\downarrow=-\downarrow\downarrow\uparrow\downarrow$  & $\uparrow\downarrow\downarrow\uparrow=-\downarrow\uparrow\uparrow\downarrow=\uparrow\uparrow\downarrow\downarrow=-\downarrow\downarrow\uparrow\uparrow$ & $\uparrow\downarrow\downarrow\uparrow=\downarrow\uparrow\uparrow\downarrow=-\uparrow\uparrow\downarrow\downarrow=-\downarrow\downarrow\uparrow\uparrow$ & $\uparrow\uparrow\downarrow\uparrow=-\downarrow\uparrow\uparrow\uparrow=-\uparrow\downarrow\downarrow\downarrow=\downarrow\downarrow\uparrow\downarrow$ \\

  &\textcolor{red}{$(-i)$ }   &  \textcolor{red}{$(-i)$ } &   & \textcolor{red}{$(-i)$ }   \\
  
\hline
$z$ & $\uparrow\uparrow\uparrow\uparrow=-\downarrow\downarrow\downarrow\downarrow=\uparrow\downarrow\uparrow\downarrow=-\downarrow\uparrow\downarrow\uparrow$ & $\uparrow\uparrow\uparrow\downarrow=\uparrow\downarrow\uparrow\uparrow=-\downarrow\uparrow\downarrow\downarrow=-\downarrow\downarrow\downarrow\uparrow$ & $\uparrow\uparrow\uparrow\downarrow=-\uparrow\downarrow\uparrow\uparrow=-\downarrow\uparrow\downarrow\downarrow=\downarrow\downarrow\downarrow\uparrow$ & $\uparrow\uparrow\uparrow\uparrow=\downarrow\downarrow\downarrow\downarrow=-\uparrow\downarrow\uparrow\downarrow=-\downarrow\uparrow\downarrow\uparrow$\\

 &   &  &\textcolor{red}{$(-i)$ }    &   \\

\hline
\hline
\end{tabular*}
\end{center}
\end{table*}

\section{Magnetic anisotropy}

The results of the previous section naturally leads to the concept of magnetic anisotropy in 
two-dimensional ferromagnets. In the present case of QHF,
Eq.~(\ref{eq:HF-energy}) applies and the possible types of magnetic anisotropies are embodied in the quadratic coefficients $U_{i,i}$.
In this section we will provide general expressions for these magnetic anisotropy coefficients $U_{ij}$ covering all possible crossings between
two LLs. To calculate the anisotropy coefficients $U_{ij}$ we will use the expressions
(\ref{Uij}), (\ref{v}), and Table~\ref{Table2}.
We begin by considering the crossing of LLs that belong to the same subband. 

\subsection{Crossing of Landau levels from the same subband: $\xi(\downarrow)=\xi(\uparrow)$}

In this case only two LLs with different spins can be aligned. From Eq.~(\ref{master}),
this means that $p_1' = p_1 = \pm \, p$, and $p_2' = p_2 = \pm \, p$. This condition arises from the first two delta
functions that impose real-spin conservation at the ``scattering'' process, while the remaining delta
function involving the orbital quantum numbers is satisfied automatically:
$n(p_1')-n(p_1) = n(\pm p)-n(\pm p) = 0 = n(\pm p)-n(\pm p) = n(p_2')-n(p_2)$.
From the same Eq.~(\ref{master}), we obtain then that

\begin{equation}
 v_{p,p,p,p}^{\parallel}(q) = \frac{2\pi}{q}\left[L_{n(p)}\left(\frac{q^{2}}{2}\right)\right]^{2} \; ,
\end{equation}

\begin{equation}
 v_{p,-p,p,-p}^{\parallel}(q) = \frac{2\pi}{q}L_{n(p)}\left(\frac{q^{2}}{2}\right)L_{n(-p)}\left(\frac{q^{2}}{2}\right),
 \label{eq:inplane_equal_s_different_n} 
\end{equation}
while for all other cases $v_{p_1',p_2',p_1,p_2}^{\parallel}(q) = 0$.
Using now Eq.~(\ref{Uij}) and after inspection of Table I, it is easy to conclude that there is only one
non-zero magnetic anisotropic term,
\begin{eqnarray}
 U_{zz} & = & -\frac{1}{4}\int_{0}^{\infty} dq \; e^{-\frac{q^{2}}{2}}
            v_{\xi,\xi,\xi,\xi}^{\perp}\left(q\right) \nonumber \\
            &  & \times \left[L_{n(p)}\left(\frac{q^{2}}{2}\right)-
            L_{n{(-p)}}\left(\frac{q^{2}}{2}\right)\right]^{2}  \; .
 \label{eq:Same_subband_Uzz}
\end{eqnarray}
If the two LLs have the same orbital quantum number ($n(p)=n(-p)$), then $U_{zz}=0$
and the ferromagnetic state at the crossing is isotropic~\cite{isotropic}. 
In the isotropic universality class, all pseudospin magnetization directions have identical energy, only the ground state has long-range order,
and there are no finite-temperature transitions.
A particularly important example of this case here is 
the $\nu=1$ situation with
$\xi(\downarrow)=\xi(\uparrow)=-1,n(\downarrow)=n(\uparrow)=0,\sigma(\downarrow) \neq \sigma(\uparrow)$.
On the other side, since $v_{\xi,\xi,\xi,\xi}^{\perp}\left(q\right)>0$,
for $n(\uparrow)\neq n(\downarrow)$ Eq.~(\ref{eq:Same_subband_Uzz})
implies that $U_{zz}<0$ and the system has a $z$ easy-axis anisotropy ($m_z=\pm1$).
In this type of universality class with discrete directions at which the energy of the ordered state is minimized, the system have long-range
order at finite temperature and phase transitions of the Ising type.
A particularly important example of this situation is the $\nu=2$ case corresponding to 
$\xi(\downarrow)=\xi(\uparrow)=-1,n(\downarrow)=0,n(\uparrow)=1,\sigma(\downarrow)=-\frac{1}{2},\sigma(\uparrow)=\frac{1}{2}$.
Here, the only difference  between the bilayer and the trilayer is the subband factor 
$v_{\xi,\xi,\xi,\xi}^{\perp}(q)$, whose sign is however the same in both cases (positive)~\cite{example}. 

\subsection{Crossing of Landau levels from different subbands: $\xi\left(\downarrow\right)\protect\neq\xi\left(\uparrow\right)$}

We will analyze now the crossing between two LLs with different
subband indices. Several cases are possible, as follows:

\subsubsection{Same spin and orbital quantum number: $\sigma(\downarrow)=\sigma(\uparrow)$,
$n(\downarrow)=n(\uparrow)$}

In this situation the three delta functions in Eq.~(\ref{master}) are satisfied in all cases,
yielding that
\begin{equation}
 v_{p_{1}',p_{2}',p_{1},p_{2}}^{\parallel}(q) =  \frac{2\pi}{q}\left[L_{n}\left(\frac{q^{2}}{2}\right)\right]^{2},
 \label{eq:inplne_equal_s_equal_n}
\end{equation}
for all possible choices of the four pseudospin indices; $L_n(x)$ are the Laguerre polynomials. Using once again
Eq.~(\ref{Uij}) and Table I, as in the previous case, one obtains now four non-zero anisotropic terms,
\begin{eqnarray}
U_{zz} &=& \frac{d}{2l_B}\left[ 
          \left(a_{\xi\left(\uparrow\right)}^{2}-a_{\xi\left(\downarrow\right)}^{2}\right)^{2}
          + \left(c_{\xi\left(\uparrow\right)}^{2}-c_{\xi\left(\downarrow\right)}^{2}\right)^{2}\right] \nonumber \\
       &-&\frac{1}{4}\int_{0}^{\infty}dq \; e^{-\frac{q^{2}}{2}} \left[L_{n}\left(\frac{q^{2}}{2}\right)\right]^{2}
       \nonumber \\
 && \left(v_{\xi(\uparrow),\xi(\uparrow)\xi(\uparrow)\xi(\uparrow)}^{\perp}+
          v_{\xi(\downarrow),\xi(\downarrow)\xi(\downarrow)\xi(\downarrow)}^{\perp}
        -2v_{\xi(\uparrow),\xi(\downarrow)\xi(\uparrow)\xi(\downarrow)}^{\perp}\right) , \nonumber \\
        \label{eq:Uzz_samen_same_s}
\end{eqnarray}
\begin{eqnarray}
 U_{xx} &=& \frac{2d}{l_B}\left(a_{\xi(\uparrow)}^{2}a_{\xi(\downarrow)}^{2}+c_{\xi(\uparrow)}^{2}c_{\xi(\downarrow)}^{2}\right)\nonumber\\
 && -\int_{0}^{\infty} dq \; e^{-\frac{q^{2}}{2}}\left[L_{n}\left(\frac{q^{2}}{2}\right)\right]^{2}
 v_{\xi\left(\uparrow\right),\xi\left(\downarrow\right),\xi\left(\uparrow\right),\xi\left(\downarrow\right)}^{\perp}(q), \nonumber \\ 
\end{eqnarray}
\begin{eqnarray}
 U_{xz} &=& U_{zx} = \frac{d}{2l_B}\left[ a_{\xi(\uparrow)}a_{\xi(\downarrow)}
 \left(a_{\xi(\uparrow)}^{2}-a_{\xi(\downarrow)}^{2}\right) \right. \nonumber \\
 && \left. +b_{\xi(\uparrow)}b_{\xi(\downarrow)}\left(b_{\xi(\uparrow)}^{2}-b_{\xi(\downarrow)}^{2}\right)
   +c_{\xi(\uparrow)}c_{\xi(\downarrow)}\left(c_{\xi(\uparrow)}^{2}-c_{\xi(\downarrow)}^{2}\right)\right]\nonumber \\
 && -\frac{1}{2}\int_{0}^{\infty} dq \; e^{-\frac{q^{2}}{2}} \left[L_{n}\left(\frac{q^{2}}{2}\right)\right]^{2} \nonumber \\
 && \left(v_{\xi(\uparrow),\xi(\uparrow),\xi(\uparrow),\xi(\downarrow)}^{\perp}(q)-
 v_{\xi(\downarrow),\xi(\downarrow)\xi(\downarrow)\xi(\uparrow)}^{\perp}(q)\right) \; . 
\end{eqnarray}

The quadratic contribution in Eq.~(\ref{eq:HF-energy}) reduces here to

\begin{equation}
 \sum_{i,j=x,y,z}U_{i,j}m_{i}m_{j} = U_{xx}m_{x}^{2}+2U_{xz}m_{x}m_{z} + U_{zz}m_{z}^{2} \;  .
 \label{anisotropic}
\end{equation}

This last equation can be minimized under the constraint $m_x^2+m_y^2+m_z^2=1$ by using a Lagrange multiplier. 
If the conditions a) $U_{xz}^2-U_{xx}U_{zz}\leq0$ and b) $U_{xx},U_{zz}>0$ are satisfied, the system will have a minimum when 
c) $U_{xx}m_x+U_{xz}m_z=0$ and  d) $U_{zz}m_z+U_{xz}m_x=0$.
Since conditions c) and d) define one plane each in the 
$m_x,m_y,m_z$ space, they are satisfied simultaneously only at the intersection line between the two planes,
which is the $m_y$ axis. Assuming that the remaining conditions are satisfied, the trilayer system exhibits in
this case $y$ easy-axis anisotropy: $m_x=m_z=0, m_y=\pm 1$.
This $y$ easy-axis configuration is new in comparison with the results for bilayer, 
since it represents two degenerate states that are fixed combinations ($\theta=\pi/2,\varphi=\pm \pi/2$) of the pseudospin 
up and down states 
with a phase factor. Note that according Eq.~(\ref{co}) each electron is then in a mixed pseudospin state in the $y$ easy-axis anisotropy class, while it is in 
a pure up or down pseudospin state in the $z$ easy-axis anisotropy class. The $y$ easy-axis anisotropy stems from the fact that the up and down pseudospin states have 
different charge distributions in each layers. Hence, the total energy can be lowered more efficiently 
by mixing the up and down pseudospin states for some values of parameters.

The case $U_{xz}^2=U_{xx}U_{zz}$ is particular, since then Eq.~(\ref{anisotropic}) may be simplified to
\begin{equation}
 \sum_{i,j=x,y,z}U_{i,j}m_{i}m_{j} = \left( \sqrt{U_{xx}}\,m_{x} + \sqrt{U_{zz}}\,m_{z} \right)^{2} \;  ,
 \label{anisotropicb}
\end{equation}
whose minimum is given by the equation $\sqrt{U_{xx}}\,m_{x} + \sqrt{U_{zz}}\,m_{z}=0$. This is the 
equation that defines a single plane in the $m_x,m_y,m_z$ space, which is however away from the $x-y$ plane.
The trilayer systems exhibit easy-plane anisotropy once again. Geometrically, for this particular case,
the two planes of the general situation collapse to a single one. 
Conditions a) and b) have been checked numerically, showing that the system can indeed present either 
$y$ easy-axis or easy-plane anisotropy. Systems with easy-plane anisotropy have a continuum of coplanar pseudospin magnetization orientations
at which the energy of the ordered state is minimized, do not have long-range order but do have Korterlitz-Thouless phase transitions at finite temperature.

\subsubsection{Same orbital quantum number and different spins: 
$\sigma(\downarrow) \neq \sigma(\uparrow)$, $n(\downarrow)=n(\uparrow)$}

In the case when the crossing is between LLs with equal orbital
quantum numbers and different spins we obtain,
through similar arguments as used in the case of the crossing between LLs
in the same subband, that the only non-zero anisotropy term is $U_{zz}$, with
the same value as in Eq.~(\ref{eq:Uzz_samen_same_s}). $U_{zz}$ is
always non negative and the crossing belongs then to the easy-plane category for $U_{zz}>0$, with the $x-y$ plane as the easy-plane.  
When $U_{zz}=0$ the quantum Hall ferromagnetism at the crossing is (fine-tuning) isotropic~\cite{isotropic}.

\subsubsection{Different orbital quantum numbers and equal spins: 
$\sigma(\downarrow)=\sigma(\uparrow)$, $n(\downarrow) \neq n(\uparrow)$}

Applying once more Eq.~(\ref{master}), since the spins of the two approaching LLs are the same,
the two delta functions acting on the possible spin values are satisfied automatically, and only the delta
function acting on the orbital quantum numbers is operative. After some inspection, one concludes that only
$v_{p,p,p,p}^{\parallel}(q)$, $v_{p,-p,p,-p}^{\parallel}(q)$, and $v_{p,-p,-p,p}^{\parallel}(q)$ are different from zero. We already have
the expressions for the first two, while
\begin{equation}
 v_{p,-p,-p,p}^{\parallel}(q) = \frac{2\pi}{q}\frac{n_{<}!}{n_{>}!} \left(\frac{q^{2}}{2}\right)^{n_{>}-n_{<}}
                        \left[L_{n_{<}}^{n_{>}-n_{<}}\left(\frac{q^{2}}{2}\right)\right]^{2} \; .
\end{equation}
From Eq.~(\ref{Uij}) and Table I we obtain only three non-vanishing quadratic anisotropic terms:
\begin{eqnarray}
U_{zz} &=& \frac{d}{2l_B}\left[ \left(a_{\xi(\uparrow)}^{2}-a_{\xi(\downarrow)}^{2}\right)^{2}+
         \left(c_{\xi(\uparrow)}^{2}-c_{\xi(\downarrow)}^{2}\right)^{2}\right] \nonumber \\
 &-&\frac{1}{4}\int_{0}^{\infty} dq \; e^{-\frac{q^{2}}{2}}
 \bigg\{\left[L_{n\left(\uparrow\right)}\left(\frac{q^{2}}{2}\right)\right]^{2}
 v_{\xi(\uparrow),\xi(\uparrow),\xi(\uparrow),\xi(\uparrow)}^{\perp}(q) \bigg. \nonumber \\
 &+&\left[L_{n\left(\downarrow\right)}\left(\frac{q^{2}}{2}\right)\right]^{2}
 v_{\xi(\downarrow),\xi(\downarrow),\xi(\downarrow),\xi(\downarrow)}^{\perp}(q) \nonumber \\
 &-& \bigg. 2L_{n\left(\downarrow\right)}\left(\frac{q^{2}}{2}\right)L_{n\left(\uparrow\right)}\left(\frac{q^{2}}{2}\right)
 v_{\xi(\uparrow),\xi(\downarrow)\xi(\uparrow),\xi(\downarrow)}^{\perp})(q) \bigg\}  \; ,  
 \label{eq:Uzz_different_n_same_s}
\end{eqnarray}
\begin{eqnarray}
 U_{xx}&=&U_{yy} = -\frac{1}{2} \frac{n_{<}!}{n_{>}!} \int_0^{\infty} dq \; e^{-\frac{q^{2}}{2}}
  \left(\frac{q^{2}}{2}\right)^{n_{>}-n_{<}} \nonumber \\
 && \left[L_{n<}^{n_{>}-n_{<}}\left(\frac{q^{2}}{2}\right)\right]^{2}
 v_{\xi(\uparrow),\xi(\downarrow)\xi(\downarrow),\xi(\uparrow)}^{\perp}(q).
 \label{eq:Uxx_different_n_same_s}
\end{eqnarray}

Using that $m_{x}^{2}+m_{y}^{2}=1-m_{z}^{2}$, the quadratic contribution
to anisotropic energy is in the form 
\begin{equation}
 \sum_{i,j=x,y,z}U_{i,j}m_{i}m_{j}=\left(U_{zz}-U_{xx}\right)m_{z}^{2}+U_{xx} \; .
 \label{eq:different_n_same_s}
\end{equation}
If $U_{zz}-U_{xx}<0$ the trilayer has easy-axis anisotropy while for $U_{zz}-U_{xx}>0$
it has easy-plane anisotropy. 
The condition $U_{zz} = U_{xx}$ defines the boundary between the two possible crossings. A numerical analysis
of this case is provided in the next section, since it corresponds to the experimental relevant case $\nu=3$.

\subsubsection{Different orbital quantum numbers and different spins: 
$\sigma(\downarrow) \neq \sigma(\uparrow)$, $n(\downarrow) \neq n(\uparrow)$}

For $n(\uparrow) \neq n(\downarrow)$ and $\sigma(\uparrow) \neq \sigma(\downarrow)$,
the only difference with the previous case is that $v_{p,-p,-p,p}^{\parallel}(q) = 0$, since real-spin
must be conserved at the crossing. The only non-zero quadratic anisotropic term is 
$U_{zz}$, as given in Eq.~(\ref{eq:Uzz_different_n_same_s}). The sign of $U_{zz}$ alone determines the type of anisotropy:
$U_{zz}>0$ induces easy-plane anisotropy, $U_{zz}<0$ induces $z$ easy-axis anisotropy. $U_{zz}=0$ corresponds
to the (fine-tuning) isotropic case~\cite{isotropic}.
A detailed numerical analysis of this case is provided in the next section, since it corresponds to the experimental relevant
case $\nu=4$.

\subsection{Numerical results for the general case}

For arbitrary values of $\varepsilon_1,\varepsilon_2,\varepsilon_3$, and $t$, the eigenvalues and
eigenvectors of Eq.~(\ref{eq:tigth_binding_matrix}) should be obtained numerically. This implies that the 
subband wave-functions of Eq.~(\ref{eq:subband WF}) must be handled also numerically. One particular
case for which the analytical solution of Eq.~(\ref{eq:tigth_binding_matrix}) is available is the 
zero-bias situation $\varepsilon_1=\varepsilon_3$. We will analyze then this case first, considering
that it is also a standard experimental situation.

\subsubsection{Trilayer at zero-bias: $\varepsilon_1 = \varepsilon_3 =0$}

The subband eigenvalues $\gamma_{\xi}$ here are given by $\gamma_0=0$, and
\begin{equation}
 \gamma_{\pm 1} = \frac{1}{2} \left\{ \varepsilon_2^* \pm \left[(\varepsilon_2^*)^2+8\right]^{1/2} \right\} \; ,
\end{equation}
with $\varepsilon_i^*=\varepsilon_i/t$. The associated normalized eigenvectors are
\begin{equation}
 a_{-1} = c_{-1} = \frac{1}{(\gamma_{-1}^2+2)^{1/2}} \; , \quad b_{-1} = \frac{\gamma_{-1}}{(\gamma_{-1}^2+2)^{1/2}} \; ,
\end{equation}
for $\xi=-1$ (subband ground-state),
\begin{equation}
 a_0 = -c_0 = \frac{1}{\sqrt{2}}, \quad b_0 = 0 \; ,
\end{equation}
for $\xi=0$ (first-excited subband state), and
\begin{equation}
 a_{1} = - c_{1} = \frac{1}{(\gamma_{1}^2+2)^{1/2}} \; , \quad b_{1} = - \frac{\gamma_{1}}{(\gamma_{1}^2+2)^{1/2}} \; ,
\end{equation}
for the last-excited state.
For $\varepsilon_2^* \gg 1$, $\gamma_{-1}(\varepsilon_2^* \gg 1) \rightarrow 0$ and 
$\gamma_{1} \rightarrow \varepsilon_2^*$. 
Accordingly, $a_{-1}(\varepsilon_2^* \gg 1) = c_{-1}(\varepsilon_2^* \gg 1)\rightarrow 1/\sqrt{2}$, 
and $b_{-1}(\varepsilon_2^* \gg 1) \rightarrow 0$; this is the (effective) zero-bias bilayer limit.
For $\varepsilon_2^* \ll -1$, $\gamma_{-1} \rightarrow \varepsilon_2^*$ and 
$\gamma_{1} \rightarrow 0$. 
Accordingly, $a_{-1}(\varepsilon_2^* \ll -1) = c_{-1}(\varepsilon_2^* \ll -1)\rightarrow 0$, 
and $b_{-1}(\varepsilon_2^* \ll -1) \rightarrow -1$; this is the (effective) zero-bias monolayer limit.
Away from these two extreme limits, the system is in the trilayer regime, with electrons
populating the three layers.

Using these expressions, the analytical evaluation of the subband potential 
$v_{\xi(p_{1}'),\xi(p_{2}'),\xi(p_{1}),\xi(p_{2})}^{\perp}({q})$ in Eq.~(\ref{eq:subband factor_giant})
is feasible, for arbitrary values of the pseudospin. As an example, we shown in Fig.~2 the $\nu=3$
magnetic anisotropy phase diagram, corresponding to the crossing between two LLs with quantum numbers
$\xi(\downarrow)=-1,n(\downarrow)=1,\sigma(\downarrow)=+\frac{1}{2}$, and 
$\xi(\uparrow)=0,n(\uparrow)=0,\sigma(\uparrow)=+\frac{1}{2}$, as displayed schematically in the inset.

\begin{figure}[h]
\begin{center}
\includegraphics[width=8.6cm]{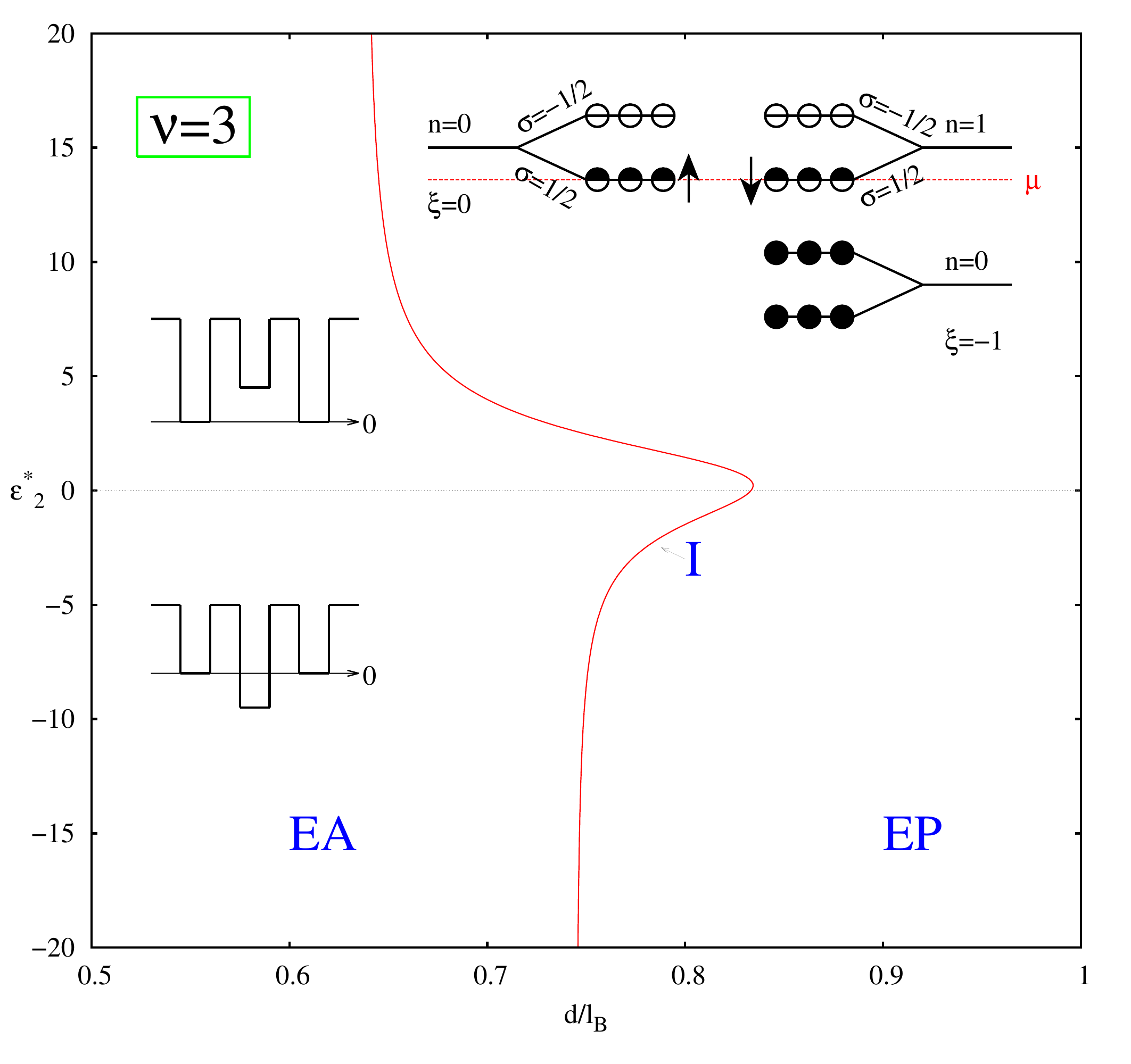}
\caption{Zero-bias $\nu=3$ magnetic anisotropy phase diagram in the $\varepsilon_2^* - d/l_B$ parameter space 
for the trilayer system. The isotropic (I) line divides the easy-axis (EA) region on the left from
the easy-plane (EP) region on the right. Left inset: schematic view of the trilayer for positive
(upper inset) or negative (lower inset) $\varepsilon_2^*$. Right inset: schematic view of the Landau
levels crossing at $\nu=3$; the crossing conserves the real-spin value. $\mu$ represents the chemical
potential. The $\uparrow,\downarrow$ symbols represent the two crossing pseudospin levels.}
\label{Fig2}
\end{center}
\end{figure}

As discussed above in Eq.~(\ref{eq:different_n_same_s}), the energetic balance dictates that the trilayer
will display EA (EP) magnetic anisotropy if $U_{zz} < (>) \, U_{xx}$, with $U_{zz}$ and $U_{xx}$ as given in
Eqs.~(\ref{eq:Uzz_different_n_same_s}) and (\ref{eq:Uxx_different_n_same_s}), respectively, particularized
for the case $\xi(\downarrow)=-1$, $\xi(\uparrow)=0$, $n(\downarrow)=1$, $n(\uparrow)=0$.
The resulting phase diagram may be understood as a competition between the Hartree energy represented by
the first term in $U_{zz}$, and the exchange contributions contained in the second term in $U_{zz}$,
and by the full $U_{xx}$.
The Hartree contribution can be evaluated analytically in this zero-bias case, yielding
\begin{eqnarray}
 U_{zz}^H &=& \frac{d}{2l_B}\left[ \left(a_{0}^{2}-a_{-1}^{2}\right)^{2}+
         \left(c_{0}^{2}-c_{-1}^{2}\right)^{2}\right] \, , \nonumber \\ 
          &=& \frac{d}{8l_B}\left(1 - \frac{4}{(\varepsilon_2^*)^2 + 8} - 
                                \frac{\varepsilon_2^*}{\sqrt{(\varepsilon_2^*)^2 + 8}} \right) \; .
 \label{Hartree}
\end{eqnarray}
As a function of $\varepsilon_2^*$, it attains its maximum value $d/4l_B$ for negative values of $\varepsilon_2^*$,
while it goes to zero for $\varepsilon_2^* \gg 1$; being always positive, it stabilizes the EP type of anisotropy.
This explains why for large enough values of $d/l_B$, the EP anisotropy dominates the phase-diagram.
On the contrary, for small values of $d/l_B$, the energetic balance is dominated by the exchange contributions,
which induces EA type of anisotropy on the left region of the phase diagram.  
For large enough $\varepsilon_2^*$, the trilayer system moves gradually to the effective bilayer configuration,
and we recover the critical value of $d/l_B$ found in Ref.(\onlinecite{JMD00}) at zero bias, after realizing that
in our effective bilayer the distance between the two occupied layers is $2d$.

The relatively large stability of the EA magnetic anisotropy for $\varepsilon_2^* \sim 0$ may be understood as follows.
Splitting $U_{zz}$ in Eq.~(\ref{eq:Uzz_different_n_same_s}) in its Hartree ($U_{zz}^H$) and exchange ($U_{zz}^X$) contributions,
the condition $U_{zz}<U_{xx}$ is rewritten as $U_{zz}^H-U_{xx}<-U_{zz}^X$; the signs here are $U_{zz}^H,-U_{zz}^X,-U_{xx}>0$.
On the other side, for increasing $\varepsilon_2^*$, $U_{zz}^H (-U_{xx})$ decreases (increases) monotonically,
while $-U_{zz}^X$ displays a weak decreasing behavior, with a shallow minimum around $\varepsilon_2^* \sim 0$.
The important point here is that $U_{zz}^H-U_{xx}$ exhibits a minimum value for $\varepsilon_2^* \sim 0$, and this
explains the reentrant behavior in the phase diagram of Fig.~3. For instance, for $d/l_B \simeq 0.8$,
by increasing $\varepsilon_2^*$ the crossing belongs to the EP anisotropy case, then to the EA type,
and finally to the EP type again.

\begin{figure}[h]
\includegraphics[width=8.6cm]{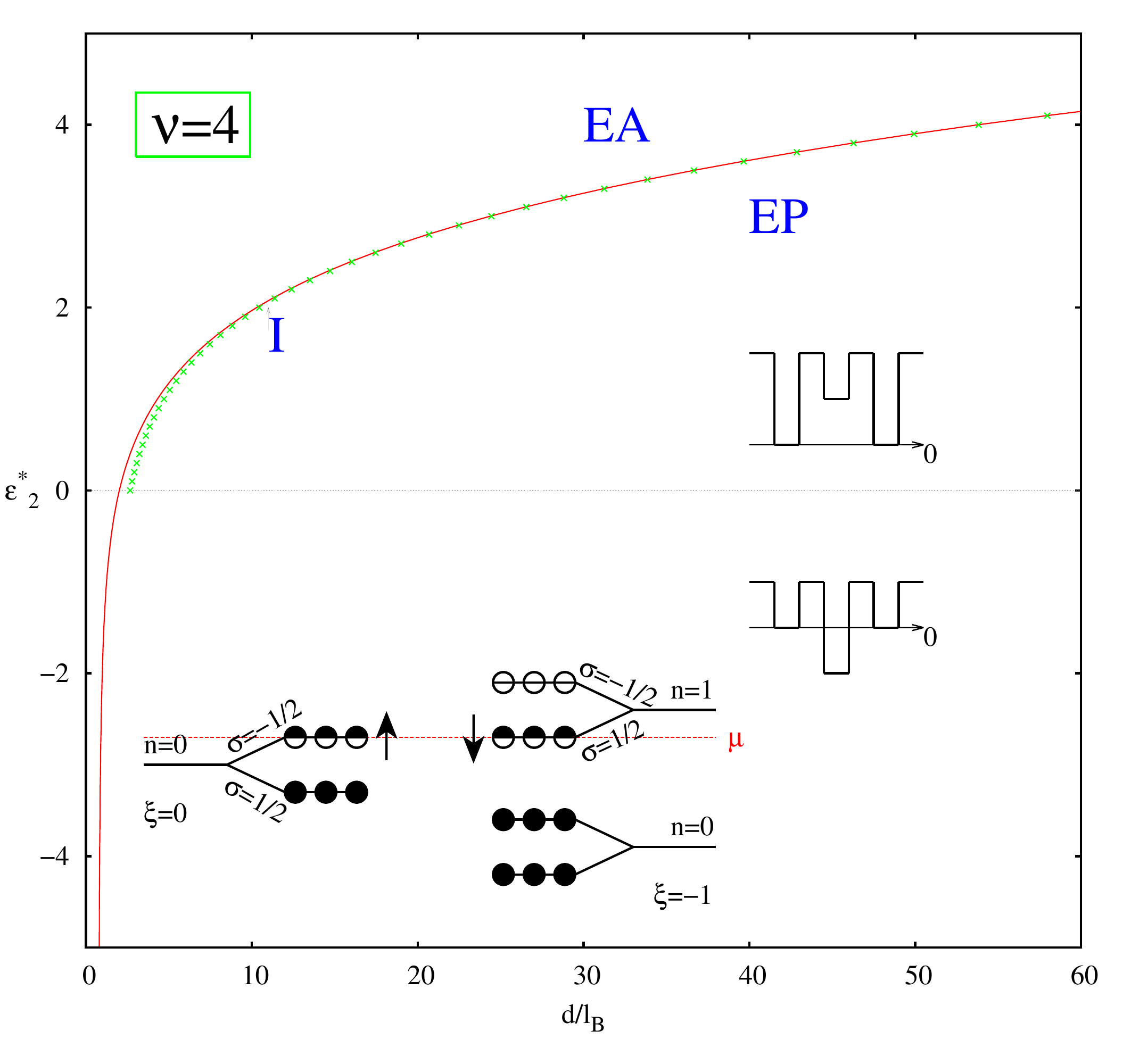}
\caption{Similar to Fig.~2, for $\nu=4$. The two crossing Landau levels have opposite real-spin values.
The dotted line corresponds to the analytical approximation of Eq.~(\ref{asymptotic}), valid for $d/l_B \gg 1$.}
\end{figure}

The magnetic anisotropy phase diagram for filling factor $\nu=4$ is displayed in Fig.~3. The two crossing
LLs have quantum numbers $\xi(\downarrow)=-1,n(\downarrow)=1,\sigma(\downarrow)=+\frac{1}{2}$, and 
$\xi(\uparrow)=0,n(\uparrow)=0,\sigma(\uparrow)=-\frac{1}{2}$, as shown in the inset. Being different the two real-spins , the 
energetic balance dictates that if $U_{zz}<(>)\,0$ the system will display EA (EP) anisotropy at the crossing situation.
The expression for $U_{zz}$ is the same as in the previous case. Once again the stability of the EP type of
anisotropy is provided by the Hartree contribution of Eq.~(\ref{Hartree}). For increasing $d$, the EP
region grows in size at the expense of the EA region. 
For large enough $d/l_B$, the boundary between the EA and the EP anisotropy phases, corresponding to the isotropic condition
$U_{zz}=0$, or $U_{zz}^H = U_{zz}^X$ may be obtained analytically

\begin{equation}
 d(\varepsilon_2^*)/l_B = \frac{\sqrt{\frac{\pi}{2}}}{8} \frac{\left(13-\frac{36}{8+(\varepsilon_2^*)^2}-\frac{7\varepsilon_2^*}{\sqrt{8+(\varepsilon_2^*)^2}}\right)}
                      {\left(1-\frac{4}{8+(\varepsilon_2^*)^2}-\frac{\varepsilon_2^*}{\sqrt{8+(\varepsilon_2^*)^2}} \right)} \, .
 \label{asymptotic}                     
\end{equation}

This equation is an explicit example of the ``fine-tuning`` of the parameters, needed to stabilize the 
isotropic type of crossing
as introduced before~\cite{isotropic}.
For large enough $\varepsilon_2^*$, we reach the 
zero-bias bilayer limit of Ref.~(\onlinecite{JMD00}), $d(\varepsilon_2^*)$ diverges, and the system displays EA anisotropy for all values of
$d/l_B$.

\begin{figure}[h]
\includegraphics[width=8.6cm]{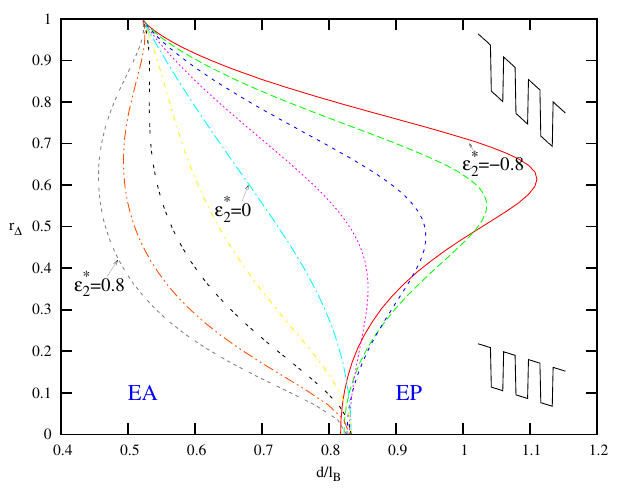}
\caption{$\nu=3$ magnetic anisotropy phase diagram in the $r_{\Delta}$-$d/l_{B}$
parameter plane, for several values of $\varepsilon_2^*$. From left to right, $\varepsilon_2^*$ = 0.8, 0.6, 0.4, 0.2, 0,
-\,0.2, -\,0.4, -\,0.6, -\,0.8. The two insets show schematically the trilayer under low ($r_{\Delta} \sim 0$) and high ($r_{\Delta} \sim 1$) bias.}
\end{figure}

\subsubsection{Trilayer with applied bias}

While the case $\varepsilon_1^*+\varepsilon_3^*=0$, $\varepsilon_2^*=0$ also allows an analytical evaluation of
Eq.~(\ref{eq:tigth_binding_matrix}) in the presence of bias, to have a more complete understanding of the possible magnetic
anisotropies we have solved numerically the case $\varepsilon_1^*=-\varepsilon_3^*=\Delta$, and arbitrary values
of $\varepsilon_2^*$. $\Delta \ne 0$ means that a bias $2 \Delta$ is applied to the trilayer (see Fig. 1).
The corresponding results are displayed in Figs.~4 ($\nu=3$) and 5 ($\nu=4$).

For discussing the trilayer properties with bias we introduce a new adimensional parameter $r_{\Delta}$, defined as
\begin{equation}
 r_{\Delta} = \frac{\Delta^*}{\sqrt{(\Delta^*)^2 + 2}} \; ,
\end{equation}
with $\Delta^*=\Delta/t$. $r_{\Delta}=0$ in the zero-bias case, while $r_{\Delta} \rightarrow \pm 1$ when $|\Delta^*| \gg 1$.

The phase-diagram for $\nu=3$ is given in Fig.~4. As before, for large enough $d/l_B$, the lineal increase
of the Hartree energy in $U_{zz}$ in Eq.~({\ref{eq:Uzz_different_n_same_s}}) with the distance between layers stabilizes the EP type of anisotropy. For small values of
$d/l_B$, on the other side, the exchange contribution in $U_{zz}$ stabilizes the EA type of anisotropy.
Interestingly, the boundary between the two phases displays a non-trivial behavior depending on the value of
$\varepsilon_2^*$. In particular, for $\varepsilon_2^* < 0$ (the standard situation in real samples), and for
$d/l_B \sim 1$, the trilayer displays first EP anisotropy at small bias, then enters in a EA regime by increasing
bias, and finally the EP anisotropy recovers by further increasing the bias. A similar situation, although somehow
less pronounced and with the role played by the EA and the EP magnetic anisotropies exchanged is observed for $\varepsilon_2^* > 0$.
By inspection of $U_{zz}^H$, one realizes that for $\varepsilon_2^* < 0$, it presents a minimum at intermediate
values of $r_{\Delta}$, that becomes deeper as $\varepsilon_2^*$ becomes more negative. For the particular
case $\varepsilon_2^*=0$, this Hartree energy may be evaluated analytically, yielding
\begin{equation}
 U_{zz}^H (r_{\Delta}) = \frac{d/l_B}{4}\frac{1+2(\Delta^*)^4}{(2+(\Delta^*)^2)^2} = \frac{d/l_B}{16}(1-2\,r_{\Delta}^2+9\,r_{\Delta}^4) \; .
\end{equation}
Minimizing this with respect to $r_{\Delta}$, one finds $r_{\Delta}^{0}=1/3$. Evaluating, one further obtains 
$U_{zz}^H(r_{\Delta}=0)= d/16l_B$, $U_{zz}^H(r_{\Delta}^{0})= d/18l_B$, and that $U_{zz}^H(r_{\Delta}=1)= d/2l_B$. 
In this case, the minimum is quite small, and the boundary between the EA and the EP phases has no evidence of a 
``reentrant'' behavior. However, as soon as $\varepsilon_2^*$ becomes more negative, we have checked numerically that the 
minimum becomes more pronounced, and moves to higher values of $r_{\Delta}$. Smaller Hartree energies help in
stabilizing the EA anisotropy, and explain why this regime increases in size for intermediate values of bias, 
as $\varepsilon_2^*$ becomes more negative. On the other side, for positive values of $\varepsilon_2^*$,
$U_{zz}^H (r_{\Delta})$ shows only a monotonic increasing behavior from $U_{zz}^H(r_{\Delta}=0)= d/16l_B$ towards
its maximum value $U_{zz}^H(r_{\Delta}=1)= d/2l_B$.

The $\nu=4$ phase diagram is displayed in Fig. 5. It shows the same reentrant behavior of the $\nu=3$ case, but 
much more enhanced: for intermediate values of $r_{\Delta}$, the EA type of anisotropy is stable for large values
of $d/l_B$, for the same moderate negative values of $\varepsilon_2^*$ as in Fig.~4. The physics is the same as in
the previous case: $U_{zz}^H(r_{\Delta})$ has a minimum at some $r_{\Delta} \neq 0 $, and stabilizes the EA anisotropy.
However, since for $\nu=4$ only the sign of $U_{zz}$ matters, the impact of this non-monotonic
behavior of $U_{zz}^H(r_{\Delta})$ on the stability of the EA anisotropy is much higher than for $\nu=3$.

\begin{figure}[h]
\includegraphics[width=8.6cm]{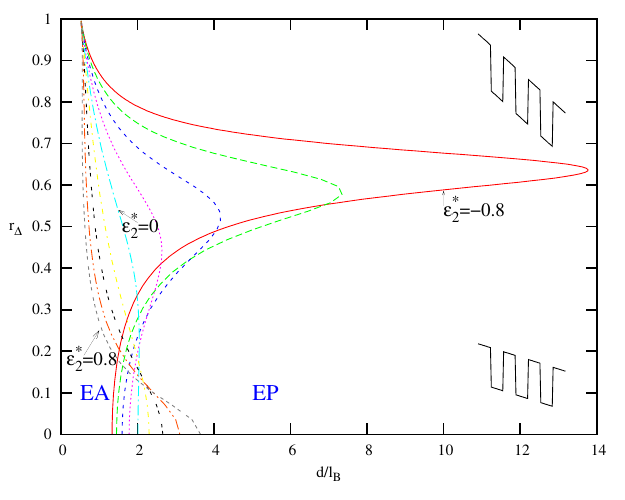}
\caption{\label{fig:nu=00003D3} $\nu=4$ magnetic anisotropy phase diagram in the $r_{\Delta}$-$d/l_{B}$
parameter plane, for several values of $\varepsilon_2^*$. From left to right, $\varepsilon_2^*$ = 0.8, 0.6, 0.4, 0.2, 0,
-\,0.2, -\,0.4, -\,0.6, -\,0.8.}
\end{figure}

By taking the appropriate limits, our present results for a trilayer reduce to the previous ones for a bilayer;
this is shown explicitly in Fig.~6, both for the $\nu=3$ and $\nu=4$ cases together. To make the comparison
more direct, we have changed the trilayer bias-related parameter $r_{\Delta}$ to
\begin{equation}
 r_{\Delta}^B = \frac{\Delta^B}{\sqrt{(\Delta^B)^2 + 4}} \; ,
\end{equation}
with $\Delta^B = \Delta^* + \varepsilon_2^*$. The idea here is that $\Delta^B$ can be small, even in the limit
$\Delta^*, -\varepsilon_2^* \gg 1$. In this limit, one of the layers will be essentially empty 
(the one with the well energy $\Delta^* \gg 1$), while the other two layers will be more or less equally occupied,
if both of them have similar well energies. In this way, we recreate the physics of the  
bilayer from the trilayer model, in the limit of large bias. 

The trilayer $\rightarrow$ bilayer evolution is easy to follow for the $\nu=3$ case. In particular,
it is quite clear how the boundary between the EA and the EP anisotropies at $r_{\Delta}^B = 0$ moves from 
trilayer values such that $d/l_B < 1$, to the bilayer critical value of $d/l_B \simeq 1.25$,
as $\varepsilon_2^*$ becomes negative.

The trilayer $\rightarrow$ bilayer evolution for the $\nu=4$ case is not so straightforward, and it should
be thought as that the ``reentrant'' behavior shown in Fig.~5 extends to arbitrary large values of $d/l_B$,
if $\varepsilon_2^*$ is negative enough. This is already observed in Fig.~5, and is reinforced in Fig.~6,
for the cases $\varepsilon_2^*=-10$ and $\varepsilon_2^*=-80$, for which the turning point of the reentrant
behavior lies far away from the limiting value $d/l_B=5$ displayed in Fig.~6.
In both cases, the trilayer $\rightarrow$ bilayer evolution displays a non-monotonic behavior.
As $\varepsilon_2^*$ becomes increasingly negative, the EA type of anisotropy becomes first more stable than 
the EP anisotropy; but after a given (large) negative value of $\varepsilon_2^*$, the EA regime looses stability
against the EP regime, finally reaching  the bilayer limit from the ``above'' side of the limiting bilayer
boundary line.

\begin{figure}[h]
\includegraphics[width=8.6cm]{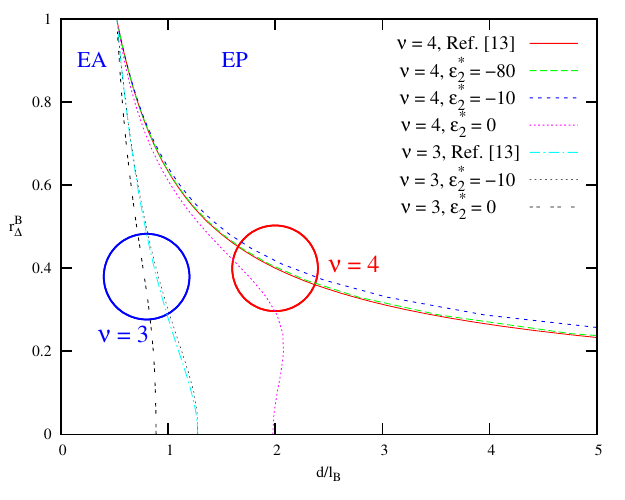}
\caption{\label{fig:Comparison-of-limit}Bilayer limit of the trilayer model, for the $\nu=3$ and $\nu=4$
crossings displayed in Figs. 3 and 4, respectively. Note that the $y$-axis ($r_{\Delta}^B$) is not the same as in these two
previous figures ($r_{\Delta}$).}
\end{figure}

\section{One-body terms}

In the previous section we have analyzed the crossing of two LLs by assuming
the absence of the linear terms in Eq.~(\ref{eq:HF-energy}). 
In the general case some of these linear terms will be present, and they will
modify the simple minimization scheme outlined in the previous section, restricted to consider only the quadratic terms in the 
pseudospin magnetization components. We will illustrate the influence of these one-body terms for the particular
cases $\nu=3$ and $\nu=4$.

At $\nu=3$ the crossing is between LLs in
different subbands, different orbital radius quantum number, and equal
spin. Then we choose $\xi=0,n=0,\sigma = 1/2$ as the pseudospin up and $\xi=-1,n=1,\sigma=1/2$
as the pseudospin down. With this in mind we write the linear terms in
Eq.~(\ref{eq:HF-energy}) as

\begin{equation}
b_{z}^*= -\frac{1}{2}\left(\gamma_{0}-\gamma_{-1}-\hbar\omega_{c}+\Delta_z^H+\Delta_z^{X}\right) \; ,
\label{eq:bz}
\end{equation}

\begin{equation}
b_{x} = \frac{1}{2} \Delta_x^H \; ,
\end{equation}
and $b_{y} = 0$.
Here $\gamma_{0}-\gamma_{-1} > 0$ is the difference between subband
energies and  $\Delta_z^X$
is the difference between the two involved pseudospin-up and pseudospin-down
exchange energies with electrons in the lower (fully occupied) $\xi=-1,n=0,\sigma=1/2$
Landau level. To calculate this exchange energy difference we can use the results of expressions
(\ref{eq:inplane_equal_s_different_n}) and (\ref{eq:inplne_equal_s_equal_n}),
obtaining 
\begin{equation}
\Delta_z^X =  \int_{0}^{\infty} dq \, e^{-\frac{q^{2}}{2}}
              \left[\frac{q^{2}}{2}v_{-1,-1,-1,-1}^{\perp}\left(q\right)-v_{0,-1,-1,0}^{\perp}\left(q\right)\right] \; .
\end{equation}
On a similar line of thought, the electrostatic (Hartree) energy imbalance  between the layers caused by
electrons in the two lower (fully occupied) Landau levels ($\xi=-1,n=0,\sigma=\pm 1/2$) are 
$V_{12}=(2d/l_B)\left(a_{-1}^{2}-b_{-1}^{2}-c_{-1}^{2}\right)$,
$V_{23}=(2d/l_B)\left(a_{-1}^{2}+b_{-1}^{2}-c_{-1}^{2}\right)$, and
$V_{13} = V_{12} + V_{23} = (4d/l_B)\left(a_{-1}^2-c_{-1}^2 \right)$. This
electrostatic (Hartree) energy may be included as effective bias ($\Delta_z^H$)
and effective tunneling ($\Delta_x^H$) parameters acting on the pseudospin
up and down states.
The resulting expressions for these Hartree contributions are
\begin{eqnarray}
 \Delta_x^H &=& \frac{8d}{l_B} \left[ a_0 \, a_{-1}(c_{-1}^2-a_{-1}^2) \right. \nonumber \\
            &-& \left. b_0 \, b_{-1}(a_{-1}^2+b_{-1}^2-c_{-1}^2)/2 \right] \; ,
\end{eqnarray}
and
\begin{eqnarray}
 \Delta_z^H &=& \frac{4d}{l_B} \left[ (a_0^2 - a_{-1}^2)(c_{-1}^2-a_{-1}^2) \right. \nonumber \\
            &-& \left. (b_0^2-b_{-1}^2)(a_{-1}^2+b_{-1}^2-c_{-1}^2)/2 \right] \; .
\end{eqnarray}
It is important to note that $\Delta_x^H$ vanishes both in the zero-tunneling and in the zero-bias limits.
In the $t=0$ case, this happens because in this limit subband and layer labels become equivalent,
and then products like $a_0 \, a_{-1}$ and $b_0 \, b_{-1}$ are just zero. In the zero-bias situation,
$a_{-1}^2=c_{-1}^2$ and $b_0=0$, resulting again in a vanishing $\Delta_x^H$. In general, this
effective tunneling parameter will be, however, different from zero, although it can be made smaller by
increasing the distance between the layers, with the associated exponential decrease of $t$~\cite{MPB16}.
Regarding $\Delta_z^H$, it is finite even in the zero-bias case, it is an even function of the bias,
and it takes its maximum value when a single layer is predominantly occupied (for instance, when 
$\varepsilon_2^* < 0$, or in the strong bias limit).

Eq.~(\ref{eq:HF-energy}) for the HF energy
also includes contributions to the lineal terms that comes from Coulomb
interactions between electrons in the two crossing pseudospin LLs. In the $\nu=3$
case only $U_{1,z}$ and $U_{z,1}$ are nonzero:

\begin{eqnarray}
 U_{1,z} & = & U_{z,1}=-\frac{1}{4}\int_{0}^{\infty}dq \, e^{-\frac{q^{2}}{2}} \nonumber \\
 &  & \times \left[v_{0,0,0,0}^{\perp}\left(q\right)-v_{-1,-1,-1,-1}^{\perp}\left(q\right)\left(1-q^{2}+\frac{q^{4}}{4}\right)\right].
 \nonumber \\
\end{eqnarray}

Then, the energy per electron (\ref{eq:HF-energy}) for the $\nu=3$ can
be written without constant terms in the form:

\begin{equation}
 e_{HF}\left(\hat{m}\right) = - \, b_x \, m_x - b_z^* \, m_{z}+\frac{1}{2}\left(U_{zz}-U_{xx}\right)m_{z}^{2} \; .
 \label{eq:total_HF_enegy_nu_3} 
\end{equation}

\begin{figure}[h]
\includegraphics[width=8.6cm]{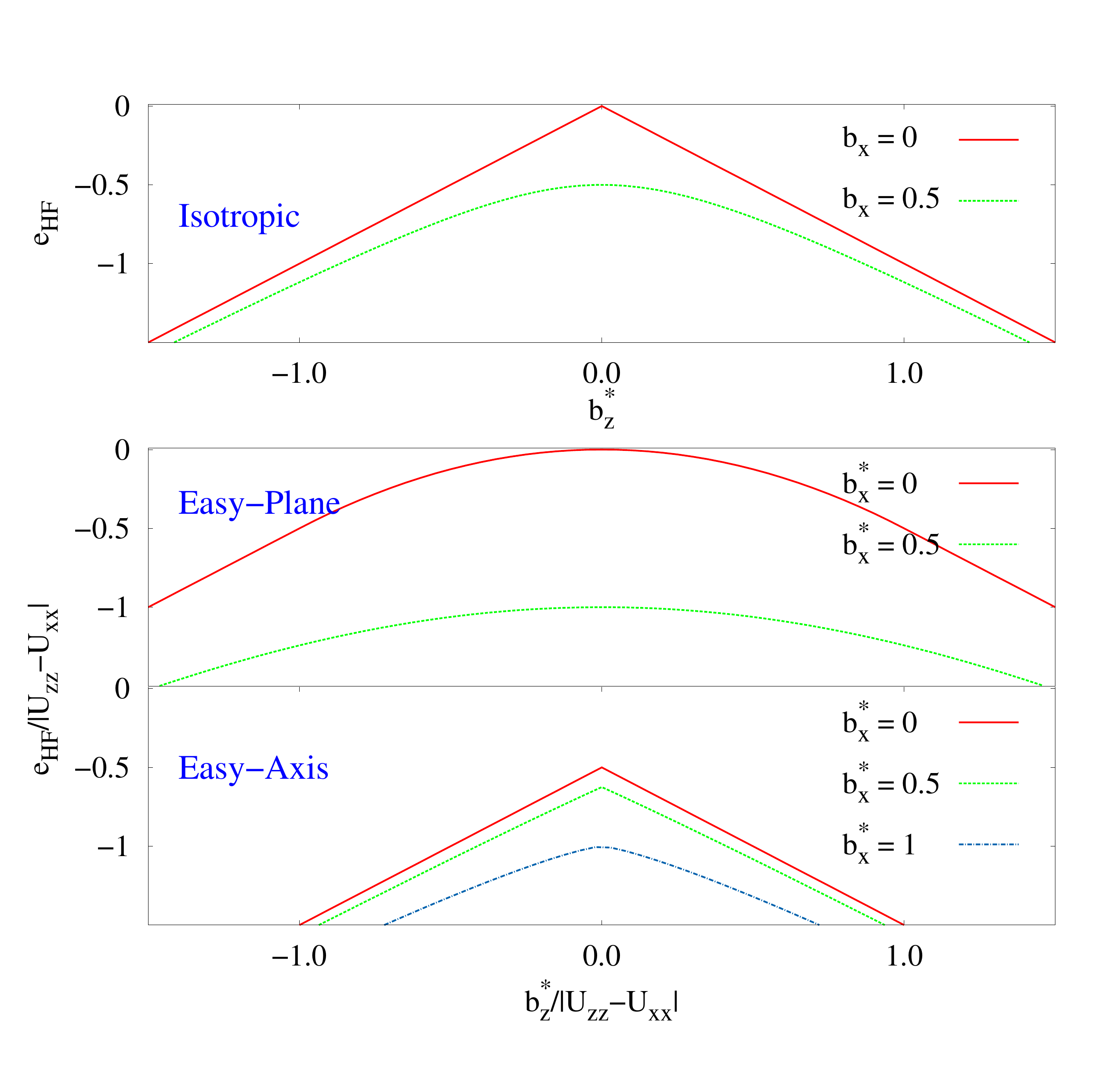}
\caption{\label{Fig7}Hartree-Fock energies as a function of the pseudospin field $b_z^*$:
upper panel, isotropic case; middle panel, easy-plane case; lower panel, easy-axis case. $b_x,b_x^*=0$
corresponds to the system with zero-tunneling between layers, while $b_z^*=0$ corresponds to the 
perfect alignment between the two crossing Landau levels. }
\end{figure}

The behavior of HF energy around the perfect alignment pseudospin field $b_z^{\ast}=0$ depends on the kind of pseudospin anisotropy. In Fig. \ref{Fig7}
we show this behavior for all possible anisotropy cases.
The upper panel corresponds to the isotropic case $U_{xx}=U_{zz}$. Eq.~(\ref{eq:total_HF_enegy_nu_3}) simplifies to 
$e_{HF}\left(\hat{m}\right) = - \, b_x \, m_x - b_z^* \, m_{z}$, and the minimization with respect to
$m_x$ and $m_z$ can be done analytically, yielding that $m_x^0 = \sign {(b_x)} \sqrt{1 / [1+(b_z^*/b_x)^2]}$,
$m_z^0 = \sign(b_z^*) \sqrt{1 / [1+(b_x/b_z^*)^2]}$, and 
\begin{equation}
 e_{HF}(m_x^0,m_z^0) = -\frac{|b_x|}{\sqrt{1+(b_z^*/b_x)^2}} - \frac{|b_z^*|}{\sqrt{1+(b_x/b_z^*)^2}} \; .
\end{equation}
In the limit $|b_z^*/b_x| \gg 1$, $e_{HF}(m_x^0,m_z^0) \rightarrow -|b_z^*|$, and note that this limit
includes the case $b_x=0$~\cite{isotropic*}. In the opposite limit $|b_z^*|/b_x \ll 1$, the HF energy displays a quadratic
behavior with the pseudospin field $b_z^*$ around the condition of perfect alignment,
$e_{HF}(m_x^0,m_z^0) \rightarrow -|b_x|-|b_z^*|^2/(2|b_x|)$. For any $b_x \neq 0$, the pseudospin rotates in the $x-z$ plane, according to the equation
$\tan \theta = \sign(b_x) \, \sign(b_z^*) \, |b_x/b_z^*|^{1/2}$, and the isotropic crossing leads 
to a smooth crossover from $m_z^0 \simeq -1$ to $m_z^0 \simeq +1$. This situation will correspond to a ``collapse'' of the 
two levels at the crossing.

The middle panel corresponds to the EP type of magnetic anisotropy ($U_{zz}-U_{xx}>0$), and in this case
the trilayer always displays a quadratic behavior in $b_z^*$, around the perfect alignment situation.
In the $b_z^* \rightarrow 0$ limit, one obtains
\begin{equation}
 \frac{e_{HF}(m_x^0,m_z^0)}{U_{zz}-U_{xx}} = -b_x^* - \frac{[b_z^*/(U_{zz}-U_{xx})]^2}{2(1+b_x^*)} \; , 
\end{equation}
with $b_x^* = b_x/|U_{zz}-U_{xx}|$. As $b_x^*$ increases, the curvature decreases, as observed in the figure.

Finally, in the lower panel the behavior of the HF energy for the EA type of anisotropy is displayed.
Interestingly, in this case the trilayer changes from having a lineal behavior with $b_z^*$ if $b_x^*$
is smaller than a critical value, to a quadratic behavior otherwise. Once again in the limit of
almost perfect alignment $b_z^* \rightarrow 0$, $m_x^0 \rightarrow b_x^*-\varepsilon$, and
$m_z^0 \rightarrow \pm \sqrt{1-(b_x^*)^2}$. Replacing in Eq.~(\ref{eq:total_HF_enegy_nu_3}) one 
obtains a lineal behavior with $b_z^*$, given approximately by 
$-b_z^* m_z^0 \sim - |b_z^*| \sqrt{1-(b_x^*)^2}$. At the critical value $b_x^* =1$, the lineal
dependence in $b_z^*$ vanishes, and it is replaced by a quadratic dependence.

At $\nu=4$  the crossing is between LLs in different subbands, different orbital quantum number, and different spin. 
Choosing $\xi=0,n=0,\sigma = -1/2$ as the pseudospin up and $\xi=-1,n=1,\sigma=1/2$ as the pseudospin down, 
we can obtain one-body terms in a way similar to the case $\nu=3$. The only difference that appears is the Zeeman term $|g|\mu_BB$ that may 
be added inside the parenthesis of Eq.~(\ref{eq:bz}). Taking this into account we obtain a similar behavior to the one 
showed in Fig.~\ref{Fig7}  for the energy  around $b_z^*=0$, after considering that in this case $U_{xx}=0$.

\begin{figure}[h]
\begin{center}
\includegraphics[width=8.6cm]{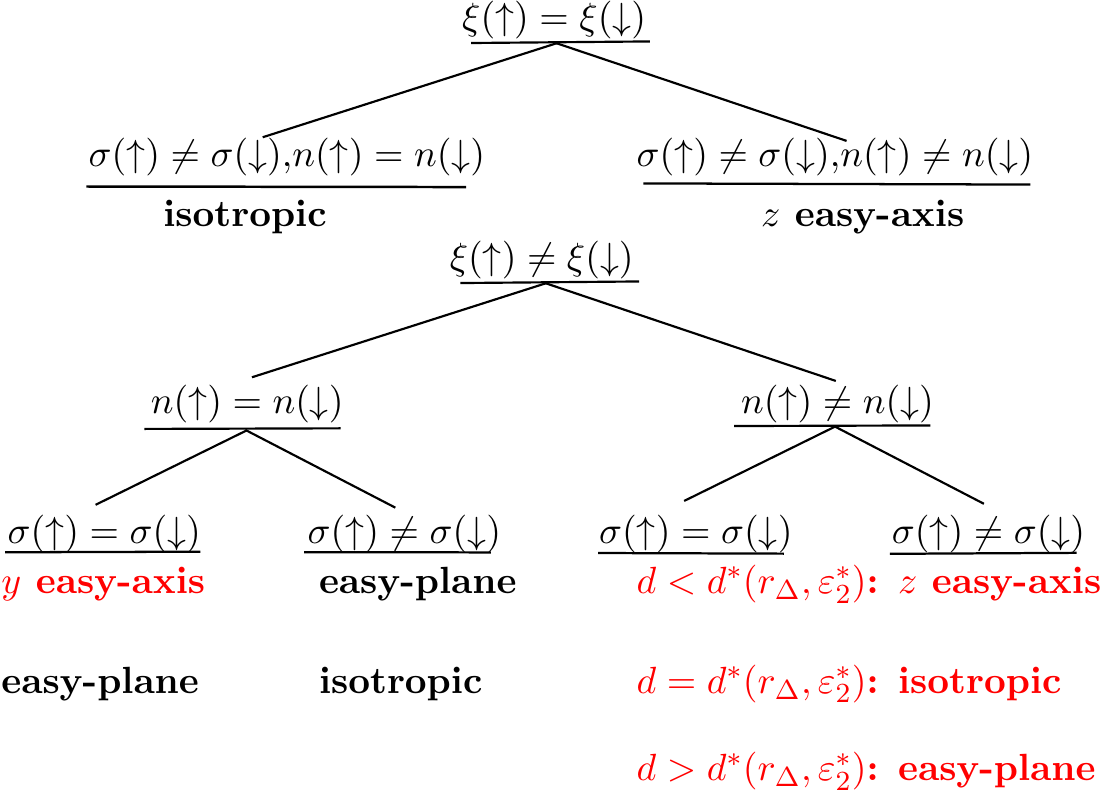}
\caption{General pseudospin magnetic anisotropy classification scheme for all possible Landau levels crossings
 in trilayers. $d^*(r_{\Delta},\varepsilon_2^*)$ denotes the boundary line between the EA and the EP magnetic anisotropies in previous figures.}
\label{Fig8}
\end{center}
\end{figure}

In Fig.~\ref{Fig8} the classifications of magnetic anisotropies obtained  in this work are schematically illustrated. 
The main differences with the bilayer cases are highlighted.

\section{Conclusions}
Using a variational Hartree-Fock approximation, we have studied theoretically all possible
two Landau levels crossings at the chemical potential, for the case of a trilayer system, in
the regime of the Integer Quantum Hall effect. The trilayer system was modeled by three
strictly bidimensional electron gases, coupled by the tunneling between neighboring layers,
and the intra- and inter-layer Coulomb interaction among electrons. The trilayer system is acted by
a strong magnetic field perpendicular to the layers, and also by an external bias that simulates
the effect of back and front gates in real experimental samples.

We have found that the general classification scheme found in bilayers also applies  to the trilayer situation,
with the simple crossing corresponding to an easy-axis ferromagnetic anisotropy analogy, and the collapse case
corresponding to an easy-plane ferromagnetic analogy. At the boundary between these two cases, an isotropic
case is also possible. While our results are valid for any filling factor $\nu$ (=1,2,3,...), we have analyzed in detail
the crossings at $\nu=3$ and $4$, and have given clear predictions that will help in their experimental search.
For instance, we have found that by increasing the bias applied to the trilayers, the system can display first easy-plane
anisotropy, then easy-axis anisotropy, and then again easy-plane anisotropy, both for the $\nu$ = 3 and 4 cases, being however the effect much more pronounced in the case $\nu=4$.
A similar reentrant behavior has also been found  at zero bias at $\nu=3$, with the energy of the central layer playing
a similar role  to the bias.

As one of the experimental techniques used in bilayers for characterizing the possible types of
magnetic anisotropies has been the measurement of the activation energies at the crossings,
we have also obtained  the zero-temperature Hartree-Fock energies close to the perfect
coincidence condition of vanishing pseudospin field, at $\nu=3$ and $\nu=4$. While the 
easy-plane HF energies always display a quadratic dependence on the pseudospin field, for the 
easy-axis anisotropy case, the trilayer at the $\nu=3,4$ crossing point exhibits a lineal dependence
on the pseudospin field, if the contribution to the Hartree energy that mixes the two pseudospins
which are in coincidence is smaller than a critical value. As this parameter is zero both in the 
zero-bias case and in the zero-tunneling case, in principle it can  be changed from one regime or
the other, either by changing the bias or doing experiments in samples with different distance between
layers. The isotropic case may have either a linear or quadratic dependence on the pseudospin field
close to the perfect alignment situation, depending on the filling factor and the quantum numbers of the 
two crossing levels.

We expect that the general classification scheme found here for all possible Landau levels crossings in semiconductor trilayers
to be as useful as the similar classification scheme of the bilayers has been.

\begin{acknowledgments}
 D.M. acknowledges the support of ANPCyT under Grant PICT-2012-0379. C.R.P. thanks
 Consejo Nacional de Investigaciones Cient\'ificas y T\'ecnicas (CONICET)
 for partial financial support and ANCyT under grant PICT-2012-0379.
\end{acknowledgments}


\begin{thebibliography}{1}
\bibitem{isotropic} The isotropic QHF is defined by the condition that no cuadratic terms in the pseudospin be present in Eq.~(\ref{HF_energy*}).
This condition may be fulfilled in several ways. For example, for the present case, this condition amounts to the constraint
$U_{zz}(d,B,\nu,t,\varepsilon_1,\varepsilon_2,\varepsilon_3)=0$, that for $\nu=1$ is satisfied automatically, without the necessity of any ``fine-tuning'' among the 
arguments of $U_{zz}$.
\bibitem{example} Note that $v_{\xi,\xi,\xi,\xi}^{\perp}(q)>0$ always, beyond the strict 2D approximation for the 
 subband wavefunctions $\lambda_{\xi}(z)$.
\bibitem{isotropic*} For $U_{zz}=U_{xx}$ and $b_x=0$, Eq.~(\ref{eq:total_HF_enegy_nu_3}) 
describes the $\nu=1$ isotropic QHF discussed above. Since the minimizing
solution is $m_z^0 = \sign(b_z^*)$, in this case the system has an abrupt change from $m_z^0=-1$
to $m_z^0=+1$, and the two pseudospin levels will just cross after coincidence.
\end{thebibliography}
\end{document}